\documentclass[12pt]{article}
\usepackage{amsmath, amssymb, color, textcomp}
\usepackage{fullpage, setspace, enumitem}
\usepackage{graphicx, subfig, epstopdf}
\usepackage{caption, titlesec, setspace, authblk, tabularx}
\usepackage{algorithm, algpseudocode, chicago}
\usepackage{tikz}
\usepackage{multicol}
\usetikzlibrary{bayesnet}

\title{Likelihood-free approximate Gibbs sampling}
\date{\today}
\titleformat*{\section}{\large\bfseries}

\author{G. S. Rodrigues\footnote{Department of Statistics, University of Bras\'{i}lia, Bras\'{i}lia, 70910-900, Brazil.}\,\,\footnote{Communicating Author: {\tt guilhermeest@yahoo.com.br}}\, ,
David J. Nott\footnote{Department of Statistics and Applied Probability, National University of Singapore, Singapore 117546.} \: 
and 
S. A. Sisson\footnote{School of Mathematics and Statistics, University of New South Wales, Sydney 2052 Australia.}}

\newcommand{\bs}[1]{\boldsymbol{#1}}
\definecolor{blue}{rgb} {0.3,0,0.8}

\newcommand{\xobs}{{\bs{X}_\mathrm{obs}}}
\newcommand{\sobs}{{\bs{s}_\mathrm{obs}}}
\newcommand{\bftheta}{\bs{\theta}}
\newcommand{\bfTheta}{\bs{\Theta}}
\newcommand{\bfmu}{\bs{\mu}}
\newcommand{\bfSigma}{\bs{\Sigma}}

\newcommand{\bflambda}{\bs{\lambda}}
       
\newcommand{\bfOmega}{\bs{\Omega}}
\newcommand{\bfepsilon}{\bs{\epsilon}}

\newcommand{\bfbeta}{\bs{\beta}}
\newcommand{\bfphi}{\bs{\phi}}
\newcommand{\bfy}{\bs{y}}

\newcommand{\bfX}{\bs{X}}
\newcommand{\bfa}{\bs{a}}
\newcommand{\bfm}{\bs{m}}
\newcommand{\bfC}{\bs{C}}
\newcommand{\bfs}{\bs{s}}
\newcommand{\bfG}{\bs{G}}
\newcommand{\bfw}{\bs{w}}
\newcommand{\bfW}{\bs{W}}
\newcommand{\bff}{\bs{f}}
\newcommand{\bfF}{\bs{F}}
\newcommand{\bfR}{\bs{R}}
\newcommand{\bfq}{\bs{q}}
\newcommand{\bfE}{\bs{E}}
\newcommand{\bfJ}{\bs{J}}
\newcommand{\bfP}{\bs{P}}

\begin{document}
  \maketitle
%   \vspace{1cm}

  \begin{abstract} 
  \noindent 
  Likelihood-free methods such as approximate Bayesian computation (ABC) have extended the reach of statistical inference to problems with computationally intractable likelihoods.
  Such approaches perform well for small-to-moderate dimensional problems, but suffer a curse of dimensionality in the number of model parameters.
  We introduce a likelihood-free approximate Gibbs sampler that naturally circumvents the dimensionality issue by focusing on lower-dimensional conditional distributions. These distributions are estimated by flexible regression models either before the sampler is run, or adaptively during sampler implementation. As a result, 
  and in comparison to Metropolis-Hastings based approaches, 
we are able to fit substantially more challenging statistical models than would otherwise be possible.
We demonstrate the
sampler's performance via two simulated examples, and a real analysis of {\em Airbnb} rental prices using a intractable high-dimensional multivariate non-linear state space model containing 13,140 parameters, which presents a real challenge to standard ABC techniques.
  \\

  \end{abstract}

  %%%%%%%%%%%%%%%%%%%%%%%%%%%%%%%
  %%%%%%%%%%%%%%%%%%%%%%%%%%%%%%%
  \section{Introduction}
  %%%%%%%%%%%%%%%%%%%%%%%%%%%%%%%
  %%%%%%%%%%%%%%%%%%%%%%%%%%%%%%%

{\em Likelihood-free} methods refer to procedures that perform likelihood-based statistical inference, but without direct evaluation of the likelihood function. This is attractive when the likelihood function is computationally prohibitive to evaluate due to dataset size or model complexity, or when the likelihood function is only known through a data generation process. Some classes of likelihood-free methods include pseudo-marginal methods \cite{Beaumont2003,Andrieu2009}, indirect inference \shortcite{Gourieroux1993} and approximate Bayesian computation \shortcite{handbook}.

In particular, approximate Bayesian computation (ABC) methods form an approximation to the computationally intractable posterior distribution by firstly sampling parameter vectors from the prior, and conditional on these, generating synthetic datasets under the model. The parameter vectors are then weighted by how well a vector of summary statistics of the synthetic datasets matches the same summary statistics of the observed data. ABC methods have seen extensive application and development over the past 15 years. See e.g.~\shortciteN{handbook} for a contemporary overview of this area.

However, ABC methods have mostly been limited to analyses with moderate numbers of parameters ($<50$) due to the inherent curse-of-dimensionality of matching larger numbers of summary statistics, in what may be viewed as a high-dimensional kernel density estimation problem \cite{blum10}. For a fixed computational budget, the quality of the ABC posterior approximation deteriorates rapidly as the number of summary statistics (which is driven by the number of model parameters) increases \shortcite{nott+ofs17}.

A number of techniques for extending ABC methods to higher dimensional models have been developed. 
Post-processing techniques aim to reduce the approximation error by adjusting samples drawn from the ABC posterior approximation in a beneficial manner. These include regression-adjustments \shortcite{Beaumont2002,blum+f10,blum+nps13},
marginal adjustment \shortcite{Nott2012},
and recalibration \shortcite{rodrigues+ps17,Prangle2013}.
However, by their nature post-processing techniques are a means to improve an existing analysis rather than a principled approach to extend ABC methods to higher dimensions. In addition, evidence is emerging that some of these procedures, in particular regression-adjustment, perform less well than is generally believed \shortcite{marin+rprr16,frazier+rr17}.

Alternative model-based approximations to the intractable posterior have been developed, including Gaussian copula models \shortcite{Li2017}, Gaussian mixture models \shortcite{bonassi+yw11}, regression density estimation \shortcite{fan+ns13}, Gaussian processes \shortcite{gutmann+c16}, Bayesian indirect inference \shortcite{Drovandi2015,drovandi+mr17}, variational Bayes \shortcite{tran+nk17} and synthetic likelihoods \shortcite{wood10,ong+ntsd16}. Each of these alternative models have appealing properties, although none of them fully address the 
high-dimensional ABC problem.

One technique that has some promise in helping extend ABC methods to higher dimensions is likelihood (or posterior) factorisation. When the likelihood can be factorised into lower dimensional components, lower dimensional comparisons of summary statistics can be made, thereby side-stepping the curse of dimensionality to some extent. This has been explored within hierarchical models by \shortciteN{Bazin2010}, within an expectation-propagation scheme by \citeN{barthelme+c14}, for discretely observed Markov models by \shortciteN{white+kp15}, and within the copula-ABC approach of \shortciteN{Li2017}.
However, such a factorisation is only available for particularly structured models (although see \shortciteNP{Li2017}).
Other approaches include rephrasing  summary statistic matching as a rare event problem \shortcite{prangle+ek16}, and using
local Bayesian optimisation techniques for high-dimensional intractable models \shortcite{meeds+w15,gutmann+c16}.

In one particular take on posterior factorisation, \shortciteN{Kousathanas2016} developed an ABC Markov chain Monte Carlo (MCMC) algorithm which only updates one parameter per iteration, so that the new candidate can be accepted or rejected based on a small subset of the summary statistics. This approach can increase MCMC acceptance rates, although it is
limited by the need to generate a synthetic dataset at each algorithm iteration, which may be computationally prohibitive if used for expensive simulators. It also requires the identification of conditionally sufficient statistics for each parameter.

In this article we introduce a likelihood-free approximate Gibbs sampler that targets the high-dimensional posterior indirectly by approximating its full conditional distributions. Low-dimensional regression-based models are constructed for each of these conditional distributions using synthetic (simulated) parameter value and summary statistic pairs, which then permit approximate Gibbs update steps. 
In contrast to 
\shortciteN{Kousathanas2016}, synthetic datasets are not generated during each sampler iteration, thereby providing efficiencies for expensive simulator models, and only require sufficient synthetic datasets to adequately construct the full conditional models (e.g.~\shortciteNP{fan+ns13}). Construction of the approximate conditional distributions can exploit known structures of the high-dimensional posterior, where available, to considerably reduce computational overheads. The models themselves can also be constructed in localised or global forms.
 
 In Section \ref{method}  we introduce the method for constructing regression-based conditional distributions and for implementing the likelihood-free approximate Gibbs sampler, and discuss possible sampler variants. In Section \ref{examples}, we explore the performance of the algorithm under various sampler and model settings, and
provide a real data analysis of an {\em Airbnb} dataset using an intractable state space model with 13,140 parameters in Section \ref{application}. Section \ref{conclusion} concludes with a discussion.

  %%%%%%%%%%%%%%%%%%%%%%%%%%%%%%%
  %%%%%%%%%%%%%%%%%%%%%%%%%%%%%%%
  \section{Likelihood-free approximate Gibbs sampler}
  %%%%%%%%%%%%%%%%%%%%%%%%%%%%%%%
  %%%%%%%%%%%%%%%%%%%%%%%%%%%%%%%
   \label{method}
   
Suppose that $\bftheta=(\theta_1,\ldots,\theta_D)^\top$ is a $D$-dimensional parameter vector, with associated prior distribution $\pi(\bftheta)$, and a computationally intractable model for data $p(\bfX|\bftheta)$. Given the observed data, $\xobs$, interest lies in the posterior distribution $\pi(\bftheta|\xobs) \propto p(\xobs|\bftheta) \pi(\bftheta)$. The ABC approximation is given by
  \begin{equation}
  \label{eqn:abcPosteriorApprox}
	\pi_\mathrm{ABC}(\bftheta|\sobs) \propto \pi(\bftheta)\int K_h(\|S(\bfX)-\sobs\|)p(\bfX|\bftheta)d\bfX,
  \end{equation}
  where $\bfs=S(\bs{X})$ is a vector of summary statistics, $\sobs=S(\xobs)$ and $K_h(u)=K(u/h)/h$ is a smoothing kernel with bandwidth parameter $h>0$. 
If the summary statistics $\bfs$ are sufficient then the approximation error can be made arbitrarily small by taking $h \rightarrow 0$ as in this case $\pi_\mathrm{ABC}(\bftheta|\sobs)$ will converge to the  posterior distribution $\pi(\bftheta|\xobs)$. Otherwise, for non-sufficient $\bfs$ and $h>0$ the approximation is given as (\ref{eqn:abcPosteriorApprox}).
See e.g.~\shortciteN{sisson+fb17} for further discussion on this approximation. A simple procedure to draw samples from $\pi_\mathrm{ABC}(\bftheta|\sobs)$ is given in Algorithm \ref{alg:basicABC}. More sophisticated algorithms are available (e.g.~\citeNP{sisson+f18}).

\begin{algorithm}[tb] 
\caption{A simple importance sampling ABC algorithm}
\label{alg:basicABC}
   
  \noindent {\it Inputs:}
  \begin{itemize}[noitemsep]
  \item An observed dataset $\xobs$.
  \item A prior $\pi(\bftheta)$ and intractable generative model $p(\bs{X}|\bftheta)$.
  \item An observed vector of summary statistics $\sobs=S(\xobs)$.
  \item A smoothing kernel $K_h(u)$ with scale parameter $h>0$.
  \item A positive integer $N$ defining the number of ABC samples.
  \end{itemize}

  \noindent {\it Data simulation and weighting:}
  
  \noindent For $i=1, \ldots, N$: 
  \begin{enumerate}[noitemsep]
  \item[1.1] Generate $\bftheta^{(i)} \sim \pi(\bftheta)$ from the prior. 
  \item[1.2] Generate $\bs{X}^{(i)} \sim p(\bs{X}|\bftheta^{(i)})$ from the model. 
  \item[1.3] Compute the summary statistics $\bfs^{(i)}=S(\bs{X}^{(i)})$.  
  \item[1.4] Compute the sample weight $w^{(i)} \propto K_h(\|\bfs^{(i)}-\sobs\|)$.
  \end{enumerate}
  
  \noindent {\it Output:} 
  \begin{itemize}
  \item A set of weighted samples $\{(\bftheta^{(i)}, w^{(i)})\}_{i=1}^N$ from $\pi_{ABC}(\bftheta|\sobs)$.
\end{itemize}
\end{algorithm}

Regression-adjustment post-processing methods \shortcite{Beaumont2002,blum+f10,blum+nps13} are commonly used to mitigate the effect of $h>0$ in (\ref{eqn:abcPosteriorApprox}) by fitting regression models of the form 
 $\theta_{d}|\bs{S} \sim f(\theta_{d} | \bs{\beta}^+_d, \bs{S})$, for $d=1, \ldots, D$, based on the weighted samples $\{(\bftheta^{(i)},\bfs^{(i)},w^{(i)})\}_{i=1}^N$, that are as close as possible to the corresponding intractable marginal distributions $\pi(\theta_{d} | \bs{S})$ in the region of $\sobs$. 
 For example, in the local linear approach of \shortciteN{Beaumont2002} the fitted models are of the form
 \[
 	\theta_d^{(i)} = \alpha_d + \bs{\beta}_d^\top(\bfs^{(i)}-\sobs) + \epsilon_d^{(i)},
\]
for $i=1,\ldots,N$ and $d=1,\ldots,D$,
where $\alpha_d\in\mathbb{R}$, $\bs{\beta}_d\in\mathbb{R}^q$, $q$ is the length of the vector of summary statistics $\bfs$, and $\epsilon_d^{(i)}\sim N(0,\sigma^2_d)$. Here $\bs{\beta}^+_d=(\alpha_d,\bs{\beta}_d,\sigma^2_d)^\top$ is the full vector of unknown regression parameters for model $d$. Regression-adjustment would then modify each $\theta_d^{(i)}$ to reduce the discrepancy between $\bfs^{(i)}$ and $\sobs$ via $\theta^{*(i)}_d=\hat{\bs{\beta}}_d^\top\sobs + (\theta_d^{(i)}-\hat{\bs{\beta}}_d^\top\bfs^{(i)})$ where $\hat{\bs{\beta}}_d$ denotes an estimated (e.g.~least squares) value of $\bs{\beta}_d$.

To construct the  likelihood-free approximate Gibbs sampler we similarly build regression models, but in this case we construct regression models of the form $\theta_d|(\bs{S},\bftheta_{-d}) \sim f(\theta_d|\bs{\beta}^+_d,g_d(\bs{S},\bftheta_{-d}))$, where $\bftheta_{-d}$ is the vector $\bftheta$ but excluding $\theta_d$, so that $f(\theta_d|\bs{\beta}^+_d,g_d(\sobs,\bftheta_{-d}))$
is as close as possible to the true conditional distribution $\pi(\theta_d|\sobs,\bftheta_{-d})$ of $\pi(\bftheta|\sobs)$. 
The functions $g_d(\bs{S},\bftheta_{-d})$ indicate the function of $\bf{S}$ and $\bftheta_{-d}$ used in the regression model to determine the conditional distribution of $\theta_d$, such as e.g.~main effects or interactions. Clearly the appropriate dependent variables will vary with $d$, but will typically be relatively low dimensional (see the analyses in Section \ref{examples} for a guide on how these may be selected).
The approximate Gibbs sampler will then cycle through each of these conditional distributions in turn, drawing $\theta_d\sim f(\theta_d|\hat{\bs{\beta}}^+_d,\sobs,\bftheta_{-d})$ for $d=1,\ldots,D$, conditioning on $\bfs=\sobs$. If  $f(\theta_d|\hat{\bs{\beta}}^+_d,\sobs,\bftheta_{-d})=\pi(\theta_d|\sobs,\bftheta_{-d})$ then the resulting Gibbs sampler will exactly target $\pi(\bftheta|\sobs)$. Otherwise, the resulting sampler will be an approximation (discussed further below).
This procedure is outlined in Algorithm \ref{alg:GibbsLocal}.

\begin{algorithm}[tb] 
\caption{Likelihood-free approximate Gibbs sampling (localised models)}
\label{alg:GibbsLocal}
   
  \noindent {\it Inputs:}
  \begin{itemize}[noitemsep]
  \item An observed dataset $\xobs$.
  \item A prior $\pi(\bftheta)$ and intractable generative model $p(\bs{X}|\bftheta)$.
  \item A sampling distribution $b(\bftheta)$ describing a region of high posterior density.
  \item An observed vector of summary statistics $\sobs=S(\xobs)$.
  \item A smoothing kernel $K_h(u)$ with scale parameter $h>0$.
  \item A positive integer $N$ defining the number of ABC samples.
  \item A positive integer $M$ defining the number of Gibbs sampler iterations.
  \item A collection of regression models $f(\theta_{d} | \bs{\beta}^+_d, g_d(\bs{S}, \bftheta_{-d}))$ to approximate each full conditional distribution $\pi(\theta_d|\sobs,\bftheta_{-d})$ for $d=1,\ldots,D$.
  \end{itemize}

  \noindent {\it Data simulation:}
  
  \noindent For $i=1, \ldots, N$: 
  \begin{enumerate}[noitemsep]
  \item[1.1] Generate $\bftheta^{(i)} \sim b(\bftheta)$ from some suitable distribution $b(\bftheta)$. 
  \item[1.2] Generate $\bs{X}^{(i)} \sim p(\bs{X}|\bftheta^{(i)})$ from the model. 
  \item[1.3] Compute the summary statistics $\bfs^{(i)}=S(\bs{X}^{(i)})$.  
  \end{enumerate}

  \noindent {\it Approximate Gibbs sampling:}  
  \begin{enumerate}
  \item[2.1] Initialise $\tilde{\bftheta}^{(0)}=(\tilde{\theta}_1^{(0)},\ldots,\tilde{\theta}_D^{(0)})^\top$.
  \item[2.2] For $m=1,\ldots,M$:\\
    \phantom{11} For $d=1,\ldots,D$:
  \begin{enumerate}
  	\item[2.2.1] Denote by $\bftheta^\star_{-d}=(\tilde{\theta}_1^{(m)},\ldots,\tilde{\theta}^{(m)}_{d-1},\tilde{\theta}^{(m-1)}_{d+1},\ldots,\tilde{\theta}^{(m-1)}_{D})^\top$ the vector containing the most recently updated values of $\tilde{\theta}^{(\cdot)}_j$, $j\neq d$.
	
  	\item[2.2.2] Set the regression weights $w_d^{(i)}=K_h(\|g_d(\bfs^{(i)}, \bftheta^{(i)}_{-d})-g_d(\sobs, \bftheta^\star_{-d})\|)\pi(\bftheta)/b(\bftheta)$ for $i=1,\ldots,N$.
      \item[2.2.3] Fit a suitable regression model \mbox{$\theta_{d}|(\bs{S}, \bftheta_{-d}) \sim f(\theta_{d} | \bs{\beta}^+_d, g_d(\bs{S}, \bftheta_{-d}))$} using the weighted samples $\{(\bftheta^{(i)},\bfs^{(i)},w_d^{(i)})\}_{i=1}^N$, so that $f(\theta_{d} | \hat{\bs{\beta}}^+_d, g_d(\sobs, \bftheta^\star_{-d}))$  locally approximates the full conditional distribution $\pi(\theta_d|\sobs,\bftheta^\star_{-d})$. 
      \item[2.2.4] Gibbs update: sample $\tilde{\theta}^{(m)}_d|(\sobs, \bftheta^\star_{-d}) \sim f (\theta_d | \hat{\bs{\beta}}^+_d, g_d(\sobs, \bftheta^\star_{-d}) )$.
  \end{enumerate}
  \end{enumerate}
  
  \noindent {\it Output:} 
  \begin{itemize}
  \item Realised Gibbs sampler output $\tilde{\bftheta}^{(0)},\ldots,\tilde{\bftheta}^{(M)}$ with target distribution $\approx\pi(\bftheta|\sobs)$.
\end{itemize}
\end{algorithm}

The algorithm begins similarly to many ABC algorithms, by drawing samples 
$\{(\bftheta^{(i)},\bfs^{(i)})\}_{i=1}^N$ from the predictive distribution $(\bftheta^{(i)},\bs{X}^{(i)})\sim p(\bs{X}|\bftheta)b(\bftheta)$ and computing $\bfs^{(i)}=S(\bs{X}^{(i)})$.
In most standard ABC algorithms $b(\bftheta)$ is the prior distribution $\pi(\bftheta)$ or an importance sampling distribution. 
Then, a standard Gibbs sampler procedure is implemented by sampling each parameter in turn from an approximation to its full conditional distribution $\theta^{(m)}_d|(\sobs, \bftheta_{-d}) \sim f (\theta_d | \hat{\bs{\beta}}^+_d, g_d(\sobs, \bftheta_{-d}) )$. These approximations are fitted using the pool of weighted samples $\{(\bftheta^{(i)},\bfs^{(i)},w_d^{(i)})\}_{i=1}^N$, where the weights $w_d^{(i)}\propto K_h(\|(\bfs^{(i)}, \bftheta^{(i)}_{-d})-(\sobs, \bftheta^\star_{-d})\|)\pi(\bftheta)/b(\bftheta)$ ensure that higher importance is given to those samples which more closely match both the observed data $\sobs$ and the conditioned values of the parameters $\bftheta_{-d}=\bftheta_{-d}^\star$. 

Clearly it is important that consideration be given to appropriate scaling of summary statistics and parameter values within the distance measure $\|\cdot\|$ to avoid one or other dominating the comparison.
Note that it is only required that the full conditionals are estimated well in regions of high posterior density, rather than over the entirety of the support of $\bftheta$. In this manner, the importance density $b(\bftheta)$ can be chosen to place $\bftheta^{(i)}$ samples in regions where the conditional distributions need to be well approximated, which may be a much smaller region than specified by the prior $\pi(\bftheta)$ (e.g.~\shortciteNP{fan+ns13}). One such strategy was successfully adopted by \shortciteN{Fearnhead2012}  who specified $b(\bftheta)$ as proportional to the prior $\pi(\bftheta)$ but restricted to a region of high posterior density as identified by a pilot simulation.

Any appropriate regression technique can be used to construct the models $f (\theta_d | \bs{\beta}^+_d, g_d(\sobs, \bftheta_{-d}) )$ such as non-parametric models, GLMs, neural networks, semi-parametric models, lasso etc. There are two possible ways to draw samples from each conditional regression model (step 2.2.4 in Algorithm \ref{alg:GibbsLocal}). The first is when a parametric error distribution has been assumed, in which case a new sample may be drawn directly from the fitted distribution. For example, if the regression model is specified such that $\theta_d\sim N(\hat{\mu},\hat{\sigma}^2)$ for specified $\hat{\mu}$ and $\hat{\sigma}^2$, then a new value of $\theta_d$ may be drawn directly from $N(\hat{\mu},\hat{\sigma}^2)$.  Alternatively, when a parametric error distribution is not assumed, the (weighted) distribution of empirical residuals $r^{(i)}_d = \theta^{(i)}_d-\hat{\mu}$ can be constructed as $R^N_d(r)=\sum_{i=1}^Nw_d^{*(i)}\delta_{r_d^{(i)}}(r)$ where $w_d^{*(i)}=w_d^{(i)}/\sum_{j=1}^Nw_d^{(j)}$, and $\delta_Z(z)$ is the Dirac measure, defined as $\delta_Z(z)=1$ if $z\in Z$ and $\delta_Z(z)=0$ otherwise. A new value of $\theta_d$ is then given by $\theta_d=\hat{\mu} + r$ where $r\sim R^N_d(r)$.

\begin{algorithm}[tb] 
\caption{Likelihood-free approximate Gibbs sampling (global models) \qquad \qquad [Changes from Algorithm 2.]}
\label{alg:GibbsGlobal}

  \noindent {\it Approximate Gibbs sampling:}  
  \begin{enumerate}
  \item[2.1] Initialise $\tilde{\bftheta}^{(0)}=(\tilde{\theta}_1^{(0)},\ldots,\tilde{\theta}_D^{(0)})^\top$.
  \item[2.2] Compute the sample weights $w^{(i)} \propto K_h(\|\bfs^{(i)}-\sobs\|)\pi(\bftheta)/b(\bftheta)$,  for $i=1,\ldots N$.
  \item[2.3] For $d=1,\ldots,D$:\\
  Fit a suitable regression model \mbox{$\theta_{d}|(\bs{S}, \bftheta_{-d}) \sim f(\theta_{d} | \bs{\beta}^+_d, g_d(\bs{S}, \bftheta_{-d}))$} using the weighted samples $\{(\bftheta^{(i)},\bfs^{(i)},w^{(i)})\}_{i=1}^N$, so that $f(\theta_{d} | \hat{\bs{\beta}}^+_d, g_d(\sobs, \bftheta_{-d}))$  locally approximates the full conditional distribution $\pi(\theta_d|\sobs,\bftheta_{-d})$.
  \item[2.4] For $m=1,\ldots,M$:\\
    \phantom{11} For $d=1,\ldots,D$:
  \begin{enumerate}
  	\item[2.4.1] Denote by $\bftheta^*_{-d}=(\tilde{\theta}_1^{(m)},\ldots,\tilde{\theta}^{(m)}_{d-1},\tilde{\theta}^{(m-1)}_{d+1},\ldots,\tilde{\theta}^{(m-1)}_{D})^\top$ the vector containing the most recently updated values of $\tilde{\theta}^{(\cdot)}_j$, $j\neq d$.

      \item[2.4.2] Gibbs update: sample $\tilde{\theta}^{(m)}_d|(\sobs, \bftheta^\star_{-d}) \sim f (\theta_d | \hat{\bs{\beta}}^+_d, g_d(\sobs, \bftheta^\star_{-d}) )$.
  \end{enumerate}
  \end{enumerate}

\end{algorithm}

%global versions. How complexity compare to other method.

The computational overheads in Algorithm \ref{alg:GibbsLocal} are in the initial data simulation stage (steps 1.1--1.3) which is standard in many ABC algorithms, and in the fitting of a separate regression model for each parameter $\theta_d$ in each stage of the Gibbs sampler (steps 2.2.2--2.2.3). For the latter, while it can be computationally cheap to fit any one regression model, repeating this $MD$ times during sampler implementation can clearly raise the computational burden. There are two approaches that can reduce these costs, which can be implemented either separately or concurrently.

In certain cases, the model $p(\bftheta|\sobs)$ will have a structure such that several of the model parameters will have exactly the same form of full conditional distribution $\pi(\theta_d|\sobs,\bftheta_{-d})$. One such example is a hierarchical model (see Section \ref{hierarchical}) where $x_{dj}\sim p(x|\theta_d)$ for $j=1,\ldots,n_d$, and $\theta_1,\ldots,\theta_{D-1}\sim N(\theta_D,\sigma^2)$. Here the form of $\pi(\theta_d|\sobs,\bftheta_{-d})$ is identical for $d=1,\ldots,D-1$. 
Accordingly the regression model $f(\theta_{d} | \bs{\beta}^+_d, g_d(\bs{S}, \bftheta_{-d}))$ can be fitted
by pooling the weighted samples $\{(\bftheta^{(i)},\bfs^{(i)},w_d^{(i)})\}_{i=1}^N$ for $d=1,\ldots,D-1$ (each using different sub-elements of the vectors), thereby allowing computational savings in allowing the value of $N$ to be reduced.
Further, in the case where the conditional independence graph structure of the posterior is known (again, consider the hierarchical model), then the choice of which elements of $\bftheta_{-d}$ should be included within the regression function $g_d(\bs{S},\bftheta_{-d})$ is immediately specified as the neighbours of $\theta_d$ on the conditional independence graph, and this does not then require independent elicitation. 
Finally, in well-structured models, some  parameters may be conditionally independent of all intractable nodes in the graph. In such cases the corresponding true conditional distribution can be directly derived, instead of approximated by a regression model (see Section \ref{application}).     

A second approach is to choose the regression model $f(\theta_{d} | \bs{\beta}^+_d, g_d(\bs{S}, \bftheta_{-d}))$ sufficiently flexibly so that not only is it a good approximation of $\pi(\theta_{d} |\sobs, \bftheta_{-d})$ when $\bftheta_{-d}$ is fixed at a particular value, $\bftheta^\star_{-d}$, within the Gibbs sampler, but that the regression model holds globally for any $\bftheta_{-d}$. Within Algorithm \ref{alg:GibbsLocal}, the approximation of $\pi(\theta_{d} |\sobs, \bftheta^\star_{-d})$ with $\bftheta_{-d}=\bftheta^\star_{-d}$ is achieved by weighting the $\{(\bftheta^{(i)},\bfs^{(i)})\}_{i=1}^N$ samples in the region of $\bftheta^\star_{-d}$ according to step 2.2.2. If the regression model $f(\theta_{d} | \bs{\beta}^+_d, g_d(\bs{S}, \bftheta_{-d}))$ was a good approximation of $\pi(\theta_{d} |\sobs, \bftheta_{-d})$ for any value of $\bftheta_{-d}$ (in the region of high posterior density), then the $\bftheta^\star_{-d}$ specific weighting of step 2.2.2 can be removed, all samples 
weighted as $w^{(i)}\propto K_h(|\bfs^{(i)}-\sobs\|)\pi(\bftheta)/b(\bftheta)$, thereby localising on summary statistics only,
and the regression models fitted once only, prior to implementing the Gibbs sampler. This global model likelihood-free approximate Gibbs sampler is described in Algorithm \ref{alg:GibbsGlobal}. Clearly the computational overheads of Algorithm \ref{alg:GibbsGlobal} are substantially lower than for the localised model version. However, the localised version may be expected to be more accurate in practice, precisely due to the localised approximation of the full conditional distributions, and the difficulty in deriving sufficiently accurate global regression models.

%what it converges to
  
In certain circumstances it can be seen that the likelihood-free approximate Gibbs sampler will exactly target the true partial posterior $\pi(\bftheta|\sobs)$. In the case where the true conditional distributions $\pi(\theta_d|\sobs,\bftheta_{-d})$ are nested within the family of distributions described by $f(\theta_{d} | \bs{\beta}^+_d, g_d(\bs{S}, \bftheta_{-d}))$, then as $N\rightarrow\infty$, which in turn allows $h\rightarrow 0$, then
\[
	f(\theta_{d} | \hat{\bs{\beta}}^+_d, g_d(\bs{S}, \bftheta_{-d}))\rightarrow \pi(\theta_d|\sobs,\bftheta_{-d})
\]
due to the law of large numbers ($N\rightarrow\infty$) and $h\rightarrow0$ eliminating the usual local ABC approximation error. In this case, then Algorithms \ref{alg:GibbsLocal} and \ref{alg:GibbsGlobal} will be exact.
In any other cases, $f(\theta_{d} | \hat{\bs{\beta}}^+_d, g_d(\bs{S}, \bftheta_{-d}))$ will be an approximation of $\pi(\theta_d|\sobs,\bftheta_{-d})$. This can be either a strong or weak approximation, whereby under a strong approximation $f(\theta_{d} | \bs{\beta}^+_d, g_d(\bs{S}, \bftheta_{-d}))$ can exactly describe $\pi(\theta_d|\sobs,\bftheta_{-d})$ but where $\hat{\bs{\beta}}^+_d$ has not converged to $\bs{\beta}^+_d$ (i.e.~finite $N$). In this case, the likelihood-free approximate Gibbs sampler comes under the noisy Monte Carlo framework of \shortciteN{alquier+feb16}.
Under a weak approximation, $\pi(\theta_d|\sobs,\bftheta_{-d})$ is not nested within the family $f(\theta_{d} | \bs{\beta}^+_d, g_d(\bs{S}, \bftheta_{-d}))$, and so $f(\theta_{d} | \hat{\bs{\beta}}^+_d, g_d(\bs{S}, \bftheta_{-d}))$ represents the closest approximation to  $\pi(\theta_d|\sobs,\bftheta_{-d})$ available within the regression model's functional constraints. This latter (weak) approximation can be arbitrarily good or poor.

When the fitted regression models only approximate the true posterior conditionals, then these may be {\em incompatible} in the sense that the set of approximate conditional distributions may not imply a joint distribution that is unique or even exists. This is equally a criticism of the ABC-MCMC sampler of \shortciteN{Kousathanas2016} as it is of the likelihood-free approximate Gibbs sampler, unless for the former  it can be guaranteed that the subset of summary statistics used to update $\theta_d$ in an ABC Metropolis-Hastings update step is sufficient for the full conditional distribution. 
See e.g.~\shortciteN{Arnold1999} for a book-length treatment of conditional specification of statistical models.
  
Incompatible conditional distributions are commonly encountered in the area of multivariate imputation by chained equations (MICE) also known as fully conditional specification (FCS), which is specifically designed for incomplete data problems \cite{Buuren2011}.
%,
In the simplified case of multivariate conditional distributions within exponential families, \shortciteN{Arnold1999} found that determining appropriate constraints on the model parameters to ensure a valid joint density was 
often unattainable.
 However, other authors have expressed uncertainty on the effects of incompatibility, and simulation studies have suggested that the problem may not be serious in practice \shortcite{Buuren2011,buuren+bgr06,drechsler+r08}.  \citeN{chen+ip15} have investigated the behaviour of the Gibbs sampler when the conditional distributions are potentially incompatible.

 However, in a more general study of parameterisation within Bayesian modelling, \citeN{gelman04} embraces the opportunities for inference based on inconsistent conditional distributions as a new class of models, motivated by computational and analytical convenience in order to bypass the limitations of joint models.
 %

  %%%%%%%%%%%%%%%%%%%%%%%%%%%%%%%
  %%%%%%%%%%%%%%%%%%%%%%%%%%%%%%%
  \section{Simulation studies} 
  %%%%%%%%%%%%%%%%%%%%%%%%%%%%%%%
  %%%%%%%%%%%%%%%%%%%%%%%%%%%%%%%
  \label{examples}

 We examine the performance of the likelihood-free approximate Gibbs sampler in two simulation studies: a Gaussian mixture model using global regression models, and in a simple hierarchical model with both local and global regression models.
  
  %%%%%%%%%%%%%%%%%%%%%%%%%%%%%%%
  \subsection{A Gaussian mixture model}
  %%%%%%%%%%%%%%%%%%%%%%%%%%%%%%%
 \label{sec:mixture}
  
We consider  the $D$-dimensional Gaussian mixture model of \shortciteN{Nott2012} where
  \begin{equation*}
   \label{likelihood} p(\bfs|\bftheta) = \sum_{b_1=0}^1 \ldots \sum_{b_D=0}^1 \left[ \prod_{i=1}^D  \omega^{1-b_i} (1-\omega)^{b_i} \right] \phi_D(\bfs|\bs{\mu}(\bs{b}, \bftheta), \bs{\Sigma}),
  \end{equation*}    
  where $\phi_D(\bs{x}|\bs{a}, \bs{B})$ denotes the multivariate Gaussian density with mean $\bs{a}$ and covariance $\bs{B}$ evaluated at $\bs{x}$, $\omega \in [0, 1]$ is a mixture weight, $\bs{\mu}(\bs{b}, \bftheta) = ((1-2b_1)\theta_1, \ldots, (1-2b_D)\theta_D)^\top$, $\bs{b}=(b_1, \ldots, b_D)^\top$ with $b_i \in \{0, 1\}$, and $\bs{\Sigma}=[\Sigma_{ij}]$ is such that $\Sigma_{ii}=1$ and $\Sigma_{ij}=\rho$ for $i \neq j$.   
For illustration we consider the $D=2$ dimensional case, with $\sobs=(5/2, 5/2)^\top$, fix $\omega=0.3$ and $\rho=0.7$ as known constants and specify $\pi(\theta_d)$ as $U(-20, 40)$ for $d=1,2$.

In this setting, the full conditional distributions for $\theta_1$ and $b_1$ are given by 
  \begin{eqnarray}
    \theta_1|(\theta_2, \bs{b}, \bfs) & \sim& N(\mu_{\theta_1}, \sqrt{1-\rho^2})I(-20<\theta_1<40),\nonumber\\     \mu_{\theta_1} 
  & =& s_1 -\rho s_2 + \rho \theta_2  -2 s_1b_1 + 2\rho s_2 b_1  -2\rho b_1 \theta_2  -2\rho \theta_2 b_2 + 4\rho b_1 b_2 \theta_2  \label{reg.theta} \\
   b_1|(\bftheta, b_2, \bfs) & \sim& \text{Bernoulli}(L(p_{b_1})),\nonumber\\
   p_{b_1} 
    & =& \ln\left(\frac{1-\omega}{\omega}\right) -  \frac{2}{1-\rho^2} s_1\theta_1 + \frac{2\rho}{1-\rho^2} s_2\theta_1 - \frac{2\rho}{1-\rho^2} \theta_1 \theta_2 + \frac{4\rho}{1-\rho^2} b_2 \theta_1 \theta_2,\nonumber
  \end{eqnarray}
  where $L(x)=1/(1+\exp(-x))$ denotes the logistic function.
The full conditional distributions for $\theta_2$ and $b_2$ may be obtained by switching the indices in the above.
 For this simple model we construct global regression models (Algorithm \ref{alg:GibbsGlobal}). We generate $N=1,000,000$ samples from the prior predictive distribution (i.e.~with $b(\bftheta)=\pi(\bftheta)$) and specify $K_h(u)$ as the uniform kernel ($h=\infty$).

 As an illustration, we first naively attempt to approximate the full conditional distribution of $\theta_1$ by a main-effects only (excluding $\bs{b}$) Gaussian regression model
  $\theta_1|(\theta_2, \bs{b}, \bfs) \sim N(\beta_0 + \beta_1 s_1 + \beta_2 s_2 + \beta_3 \theta_2, \sigma^2)$. 
 The resulting MLEs were 
 $\hat{\bs{\beta}} = (8.76, -0.31, 0, 0)^\top$ (s.e.~$=(0.019,0.001,0.001,0.001)$) and $\hat{\sigma}=16.16$,
 %.
 which suggests that $\theta_1$ is conditionally independent of $s_2$ and $\theta_2$. This can clearly be seen to be incorrect based on a simple graphical exploration of the synthetic samples. This is a clear warning of the need to consider sufficiently flexible regression models, with interaction effects (as discussed in \shortciteNP{Nott2012} and as is evident in the form of $\mu_{\theta_1}$).
Instead, we specify the regression mean with all main effects and interactions and, because the number of samples $N$ is large, the resulting MLEs of $\bs{\beta}$ (and $\sigma^2$) matched the true values in (\ref{reg.theta}) up to at least one decimal place (not shown).

\begin{figure}[tbh]
	\centering
	\subfloat[\footnotesize{KDE of diff. between fitted and true means}\label{reg1a}]{\includegraphics[width=5.5cm,height=7.5cm,angle=-90]{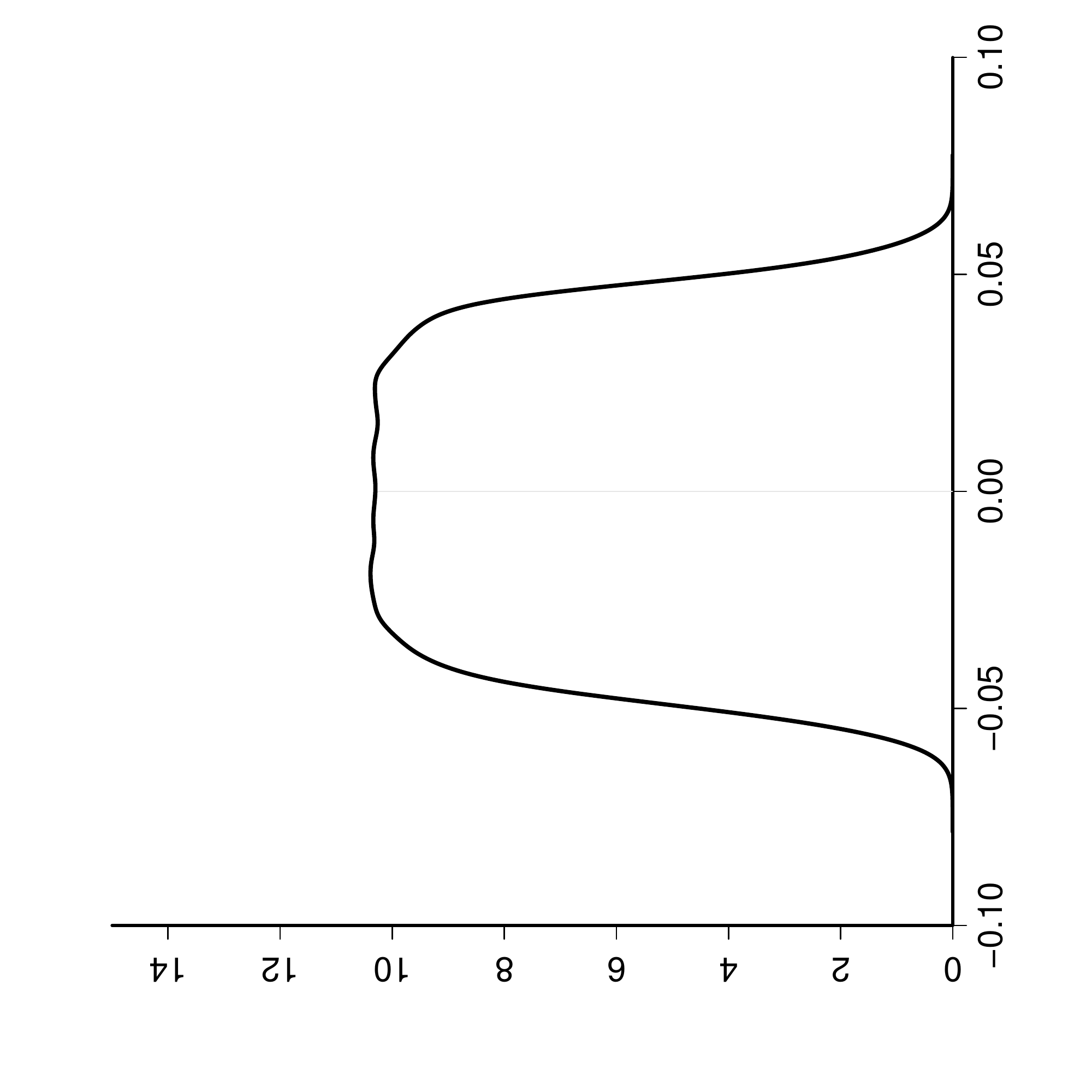}} 
	\subfloat[\footnotesize{Residual error distribution}\label{reg1b}]{\includegraphics[width=5.5cm,height=7.5cm,angle=-90]{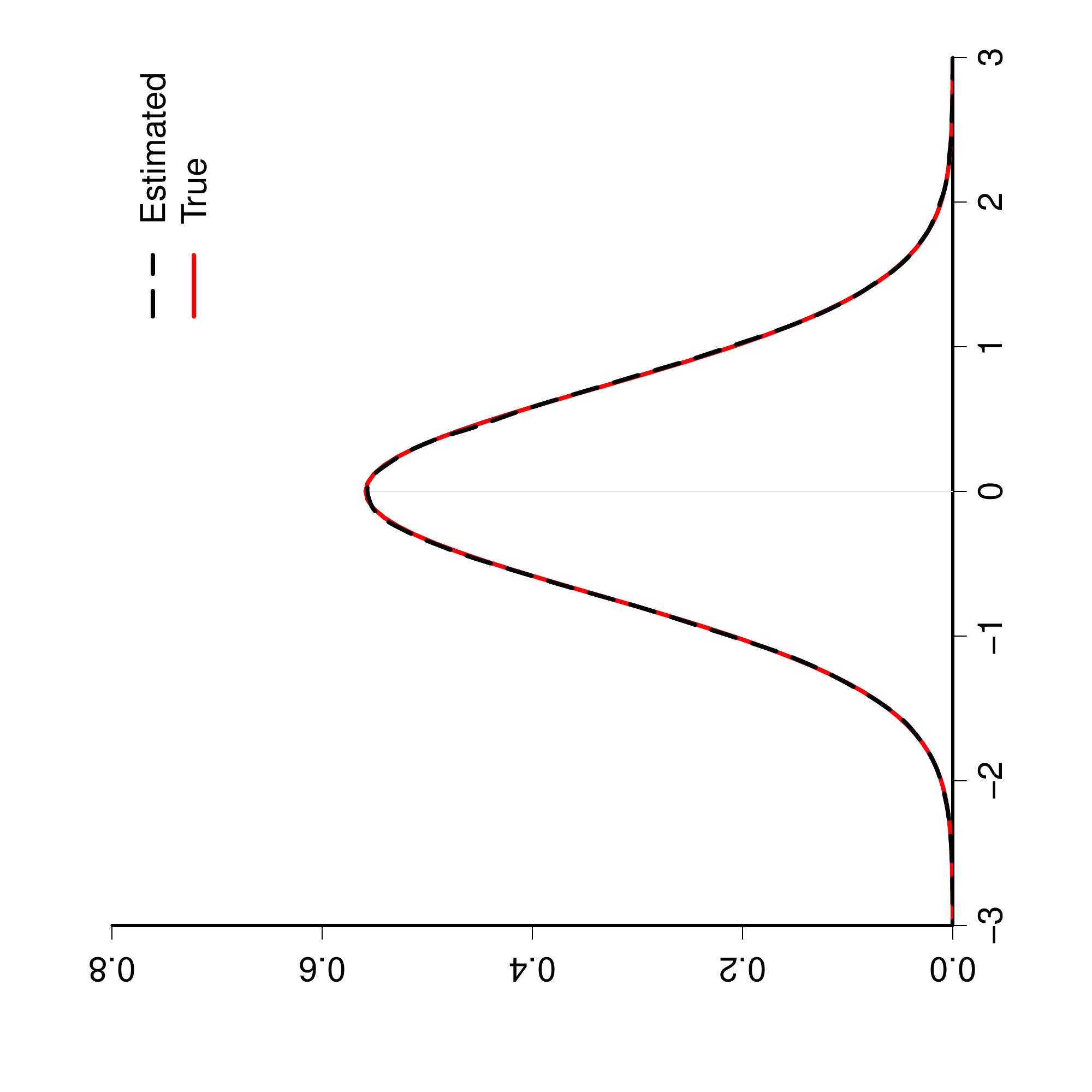}}
	\\
	\subfloat[\footnotesize{$p(b_1=1|\theta_1, s_1=s_2=2.5, b_2=0, \theta_2=-2.5)$}\label{reg1c}]{\includegraphics[width=5.5cm,height=7.5cm,angle=-90]{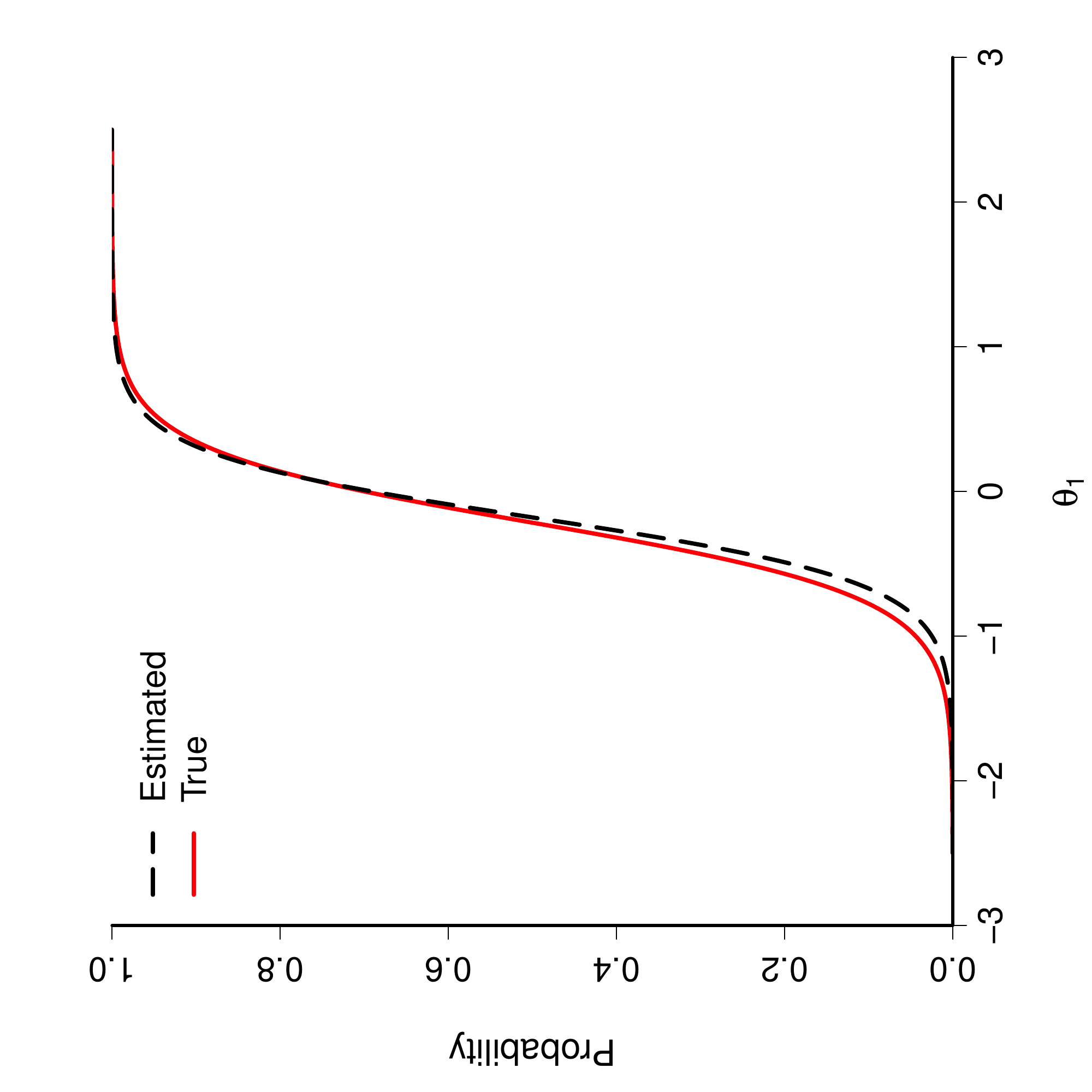}}
	\subfloat[Probability of changing the state of $b_1$ \label{reg1d}]{\includegraphics[width=5.5cm,height=7.5cm,angle=-90]{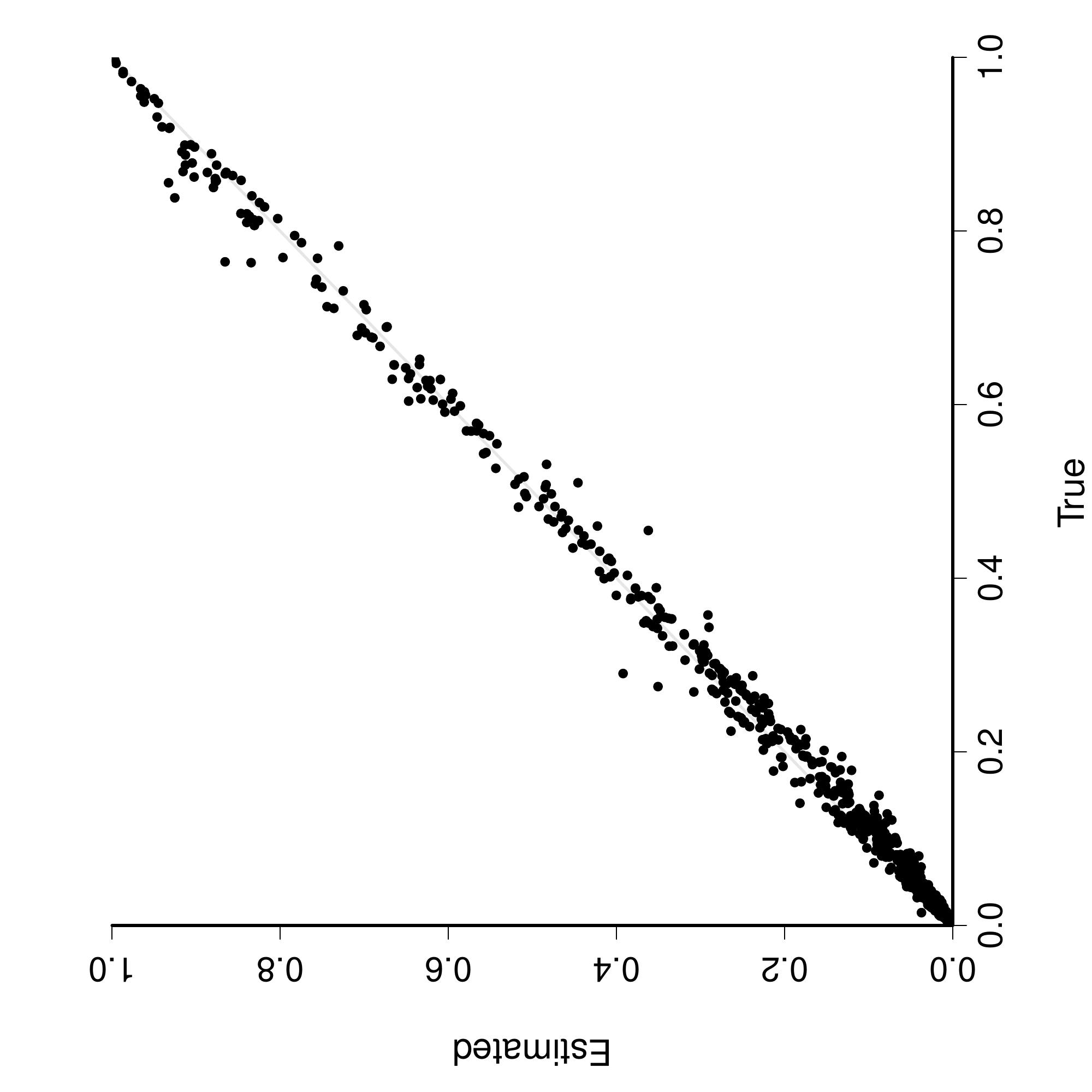}}
	\caption{\small Assessing the quality of the regression approximation. (a) A kernel density estimate (KDE) of the differences between the fitted  and true conditional mean values $\hat{\mu}^{(i)}_{\theta_1}-\mu^{(i)}_{\theta_1}$. (b) The true $N(0, \sqrt{1-\rho^2})$ error density and the KDE of the fitted regression residuals. (c) The fitted and the true conditional distribution $p(b_1=1|\theta_1, s_1=s_2=2.5, b_2=0, \theta_2=-2.5)$. (d) True versus estimated probability of changing the the state of the cluster indicator variable $b_1$.}
	\label{regression}
      \end{figure}

  Figure \ref{reg1a} shows a kernel density estimate (KDE) of the differences between the fitted and true conditional mean values ($\hat{\mu}^{(i)}_{\theta_1}-\mu^{(i)}_{\theta_1}$) for each of the $N$ data points used in the regression. In most cases, the absolute difference was less than $0.05$. Figure \ref{reg1b} shows a KDE of the empirical residuals and the true $N(0, \sqrt{1-\rho^2})$ error density. The similarity suggests that in sampling from the regression model, randomly choosing a residual is essentially equivalent to sampling from the true Gaussian error distribution.
Given that we are fitting a regression model in the same family as the true conditional distribution, we have a strong approximation of $\theta_1|(\theta_2,\bs{b},\bfs)$ (as defined in Section \ref{method}) in this case.
  
In a similar manner, we naturally model the conditional distribution of $b_1|(\bftheta,b_2,\bfs)$ as a Bernoulli GLM with logistic link function, and all possible conditional main effects and interactions.
Figure \ref{reg1c} examines the quality of this approximation by presenting the cdf's of the fitted and the true probabilities of $p(b_1=1|\theta_1, s_1=s_2=2.5, b_2=0, \theta_2=-2.5)$. The distributions are very similar, though still distinguishable. An explanation for this is that for most of the $N$ samples, the conditional probability of $b_1$ is either (numerically) 0 or 1. In other words, only the samples such that $\bftheta$ is close to the origin are informative for the regression parameters.   
This regression model is again a strong approximation to the true conditional distribution.

     \begin{figure}[htp]
	\centering
	\subfloat[\footnotesize{First 150 Gibbs iterations}\label{gibbs}]{\includegraphics[width=7.5cm,height=7.5cm,angle=-90]{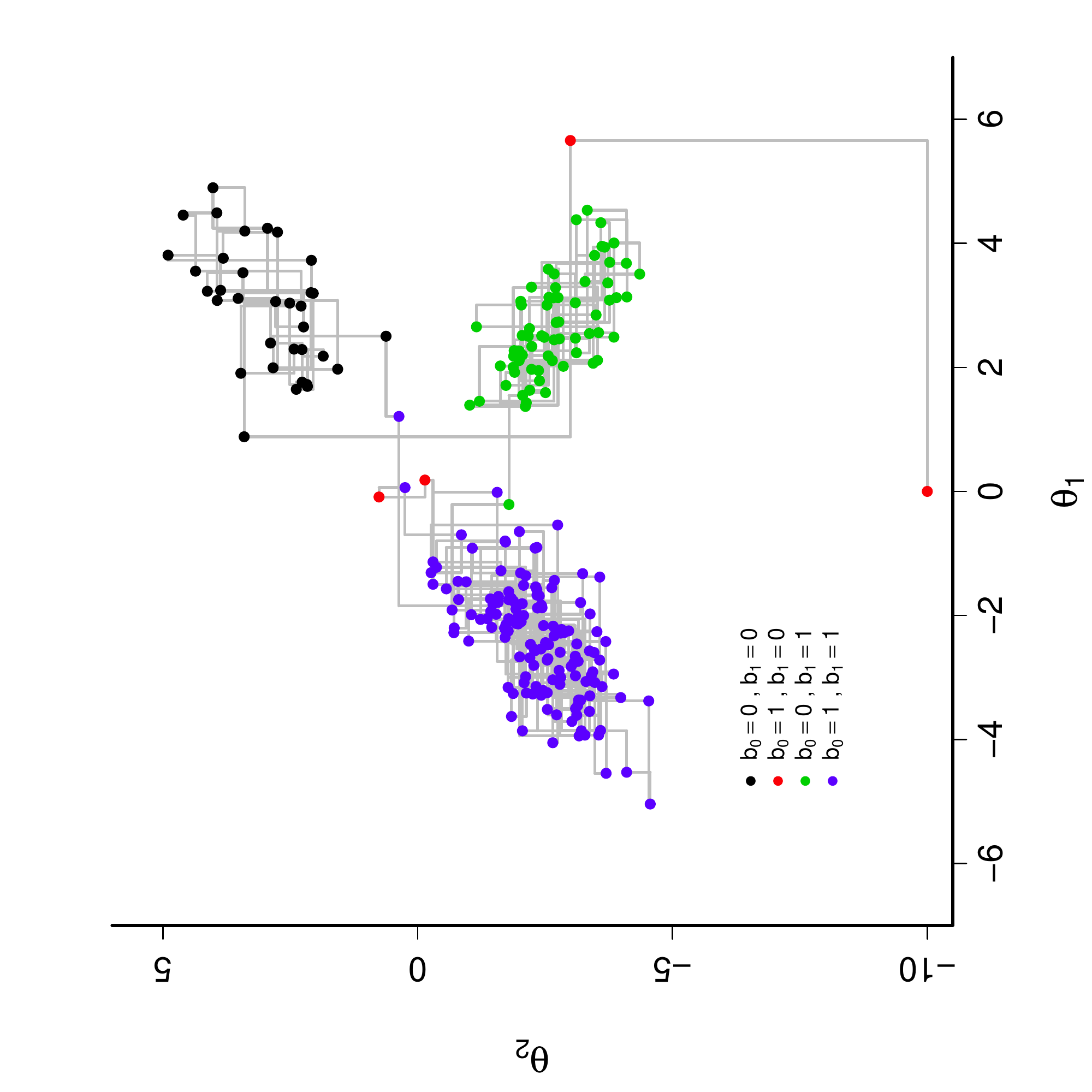}} 
	\subfloat[\footnotesize{Posterior distribution}\label{posterior}]%{\includegraphics[width=7.5cm,height=7.5cm,angle=-90]{posterior}}
	{\includegraphics[width=7.5cm,height=7.5cm,angle=-90]{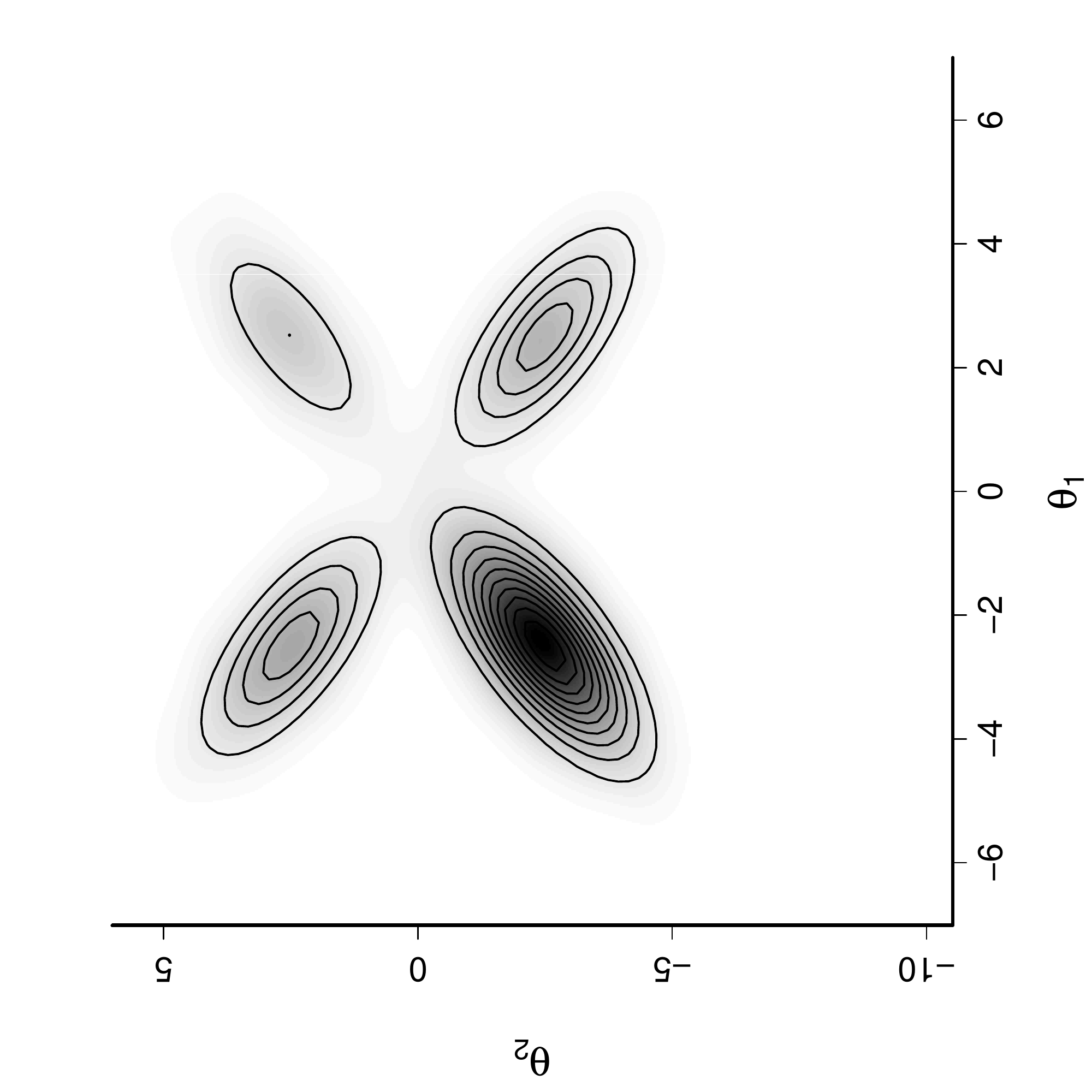}}
	\caption{\small Likelihood-free approximate Gibbs sampler output. (a) Sample path of the first 250 iterations of $(\theta_1,\theta_2)$, with the values of $(b_1,b_2)$ indicated by coloured points. (b) Posterior density estimates (shading) based on 20,000 sampler iterations, and true posterior density contours.}
	\label{results1}
      \end{figure}
      
Figure \ref{results1} illustrates the output of $M=20,000$ iterations of the resulting likelihood-free approximate Gibbs sampler, when initialised at $(\theta_1, \theta_2, b_1, b_2)=(0,-10,1,0)$. The sampler moves around the parameter space well, and visually appears to target the true posterior distribution. 
During sampler implementation the true and estimated probabilities of switching the value of $b_1$ were recorded, and are illustrated in Figure \ref{reg1d}. Only a small proportion of the $M$ probabilities are larger  than 0.2 (due to the form of the posterior), but on the whole the estimated probabilities are generally accurate, with a few exceptions.
For this example, the estimated conditional distributions (and associated switching probabilities of $b_1$) will approximate their true counterparts arbitrarily well as $N$ gets large, essentially due to the simple form of the true posterior distribution.
A better mixing approximate Gibbs sampler could also have been constructed for this posterior distribution, using a 4-level multinomial regression for the full conditional of $\bs{b}|(\bftheta,\bfs)$ and a bivariate Gaussian regression model for $\bftheta|(\bs{b},\bfs)$.

  %%%%%%%%%%%%%%%%%%%%%%%%%%%%%%%
  \subsection{A simple hierarchical model}
  %%%%%%%%%%%%%%%%%%%%%%%%%%%%%%%
\label{hierarchical}
  
We now compare the performance of a collection of approximate Gibbs sampler implementations for estimating a Gaussian hierarchical model with the ABC-MCMC method (ABC-PaSS; ABC with Parameter Specific Statistics) method introduced by \shortciteN{Kousathanas2016}, and the exact Gibbs sampler.
Hierarchical methods have been previously considered in the likelihood-free framework by e.g.~\shortciteN{Bazin2010} and \shortciteN{rodrigues+ns16}. The Gaussian hierarchical model, with parameters $\bftheta=(\mu_1, \ldots, \mu_U,  \mu, \tau_{\mu},\tau_x)^\top$, is defined as 

\vspace{.2cm}
\begin{minipage}[c]{0.55\textwidth}
\centering
\begin{align}
   \label{likelihood.hierarchical} \notag X_{u\ell} & \sim N(\mu_u, \tau^{-1}_x)  \\ 
   \notag \mu_u & \sim N(\mu, \tau^{-1}_\mu)  \\
   \notag \tau_x & \sim \text{Gamma}(\alpha_x, \nu_x) \\
   \notag \tau_\mu & \sim \text{Gamma}(\alpha_{\mu}, \nu_{\mu}) \\
   \notag \mu & \sim N(0, 1),
\end{align}
\vspace{-1cm}
\end{minipage}
\begin{minipage}[c]{0.45\textwidth}
   \tikz{ %
    \node[latent] (mu) {$\mu$} ; %
    \node[latent, below=of mu, yshift=.25cm] (mu_u) {$\mu_u$} ; %
    \node[latent, right=1 of mu_u] (tau_mu) {$\tau_\mu$} ; %
    \node[obs, below=of mu_u, yshift=.25cm] (x) {$X_{u\ell}$} ; %
    \node[latent, right=of x, xshift=0cm] (tau_x) {$\tau_x$} ; %
    \plate[inner sep=0.3cm,xshift=0cm,yshift=0cm] {plate1} {(x)} {\scriptsize $\ell=1, \ldots, L$}; %
    \plate[inner sep=0.3cm, xshift=-0cm, yshift=0cm] {plate2} {(mu_u) (plate1)} {\scriptsize $u=1, \ldots, U$ \phantom{a}}; %
    \edge {mu, tau_mu} {mu_u} ; %
    \edge {mu_u, tau_x} {x} ; %
    }
\end{minipage}
  
\vspace{1.2cm}
\noindent where $X_{u\ell}$ denotes the $\ell$-th observation in group $u$, for $\ell=1, \ldots, L$ and $u=1, \ldots, U$.
%,
The model is tractable, allowing direct comparison between the exact and approximate posteriors. 
The full conditional distributions and prior specification for this model are given in Table \ref{tab:True.FCDS}.
  
{\setlength{\tabcolsep}{1.0em}
\bgroup
\def\arraystretch{1.7}%
  \begin{table}[]
    \begin{tabular}{c|l|l|c|c}
     & Prior                                     & Full conditional distribution & Estimate? & Summary   \\ \hline
    $\mu$       & $N(0,1)$                 & $N \left( \frac{U \tau_\mu \overline{\mu}}{1 + U \tau_\mu}, (1 + U \tau_\mu)^{-1} \right)$  & $\times$  & -- \\ \hline
     $\tau_\mu$  & $\text{Ga}(\alpha_{\mu}, \nu_{\mu})$ & $\text{Ga} \left( \alpha_\mu + \frac{U}{2}, \;
  \nu_\mu + \frac{\sum_{u=1}^U (\mu_{u} - \mu)^2}{2} \right)$    & $\times$   & -- \\ \hline
    $\tau_x$    & $\text{Ga}(\alpha_x, \nu_x)$         &  $\text{Ga} \left( \alpha_x + \frac{U L}{2}, \; 
  \nu_x + \frac{\sum_{u=1}^U \sum_{\ell=1}^L (X_{u\ell} - \mu_u)^2}{2} \right)$   & \checkmark  & $\bs{S}_{\tau}$  \\ \hline
    $\mu_u$     & $N(\mu, \tau^{-1}_\mu)$       &  $N \left( \frac{\mu \tau_\mu + L \tau_x \overline{X}_u}
    {\tau_\mu + L \tau_x}, \; 
    (\tau_\mu + L \tau_x)^{-1} \right)$ & \checkmark & $\bs{S}_{u}$\\
\end{tabular}
\caption{{\small Prior and full conditional distributions of the Gaussian hierarchical model. Non-estimated conditionals ($\times$) use the full conditional distribution within the Gibbs/ABC-PaSS samplers.
}}
\label{tab:True.FCDS}
\end{table}
\egroup

    The structure of this model may be exploited to simplify sampler computations in three meaningful ways, as discussed in Section \ref{method}. 
  First, $\pi(\mu_u|\sobs, \bftheta_{-u})$ is identical for $u=1,\ldots,U$, so these distributions only need to be approximated for one group. 
  Second, the nodes which should be included within the regression function $g_d(\bs{S},\bftheta_{-d})$ are easily identified from the graph. 
  Third, it is only necessary to approximate the full conditional distribution of parameters that are conditionally dependent on intractable quantities. In the following we only update $\mu_u$ and $\tau_x$ using approximate likelihood-free methods, and use the full conditional distributions for $\mu$ and $\tau_\mu$ (Table \ref{tab:True.FCDS}).

  We compare the exact Gibbs sampler with three different approximation strategies 
  (each using the same $g_d(\cdot)$ functions):
   (a) {\em Simple global:} The conditional models are approximated by global linear regression models (Algorithm \ref{alg:GibbsGlobal});  
   (b) {\em Simple local:} The conditional models have the same linear form as the simple global approach, but fits are localised at each Gibbs iteration (Algorithm \ref{alg:GibbsLocal});
   (c) {\em Flexible global:} The conditional models are globally approximated (Algorithm \ref{alg:GibbsGlobal}) by non-linear conditional heteroscedastic feed-forward multilayer artificial neural network models (\shortciteNP{blum+f10}).

The distribution of each unit mean $\mu_1, \ldots, \mu_U$ depends on the data exclusively through the corresponding unit-specific summary statistics $\bs{S}_u=( \overline{X}_u, \hat{\tau}_u)^\top$ (e.g.~\shortciteNP{Bazin2010}), where $\overline{X}_{u}$ and $\hat{\tau}_u$ are the sample mean and precision of the data in group $u$, respectively, and therefore we take $g_{u}(\bs{S}, \bftheta_{-u})=(1, \mu, \tau_\mu, \tau_x, \bs{S}_u^\top)^\top$.
Recall (Table \ref{tab:True.FCDS}) that the conditional mean $\mathbb{E}(\mu_u|\ldots)$ is a non-linear function of the covariates, and the conditional variance $\mathbb{V}(\mu_u|\ldots)$ is not constant throughout the covariate space. 
Consequently,
for the linear and non-linear model approaches, we approximate the true conditional distribution by
\begin{eqnarray*}
\mu_u | (\bs{S}, \bftheta_{-u}) &\sim& N \left((1, \mu, \tau_\mu, \tau_x, \bs{S}_u^\top)^\top \bs{\beta}_{\mu_u}, V_{\mu_u} \right),\\
\mu_u | (\bs{S}, \bftheta_{-u}) &=& m(g_{\mu_u}(\cdot)) + \sigma(g_{\mu_u}(\cdot)) \times \zeta,
\end{eqnarray*}
respectively,
where $\zeta$ is a random variable with mean zero and fixed variance. 
The conditional expectation is estimated as $\hat{m}(g_{\mu_u}(\cdot))$ with a neural network using the 
 \emph{R} function {\em h2o.deeplearning} \shortcite{H2o2018} with default model settings. The variance term $\sigma(g_{\mu_u}(\cdot))$ is similarly estimated by a gamma neural network fitted over the squared residuals $r^2=({\mu_u} - \hat{m}(g_{\mu_u}(\cdot)))^2$. 
An approximate sample from the full conditional distribution is then
\[
 {\mu_u^*}=\hat{m}(g_{\mu_u}(\cdot)) + \hat{\sigma}(g_{\mu_u}(\cdot)) \times \frac{{\mu_u}^{(i)} - \hat{m}^{(i)}}{\hat{\sigma}^{(i)}},
 \]
where $i$ is randomly selected from $1, \ldots, N$ (step 2.2.4 in Algorithm \ref{alg:GibbsLocal}).

For the full conditional distribution of $\tau_x$, after discarding uninformative nodes, we defined 
\[
g_{\tau_x}(1, \mu_1, \ldots, \mu_U, \bs{S}_\tau)=
(1, \overline{\mu}, \hat{\tau}_{\mu_u}, \bs{S}_\tau^\top)^\top,
\qquad
\bs{S}_\tau=(\overline{\overline{X}}, \tau_{\overline{X}}, \overline{\hat{\tau}}, \tau_{\hat{\tau}})^\top
\]
where the {\em symmetric} summary statistics are $\overline{\overline{X}} = \frac{1}{U}\sum_{u=1}^U \overline{X}_{u}$, $\overline{\hat{\tau}} = \frac{1}{U}\sum_{u=1}^U \hat{\tau}_u$, 
$$
\tau_{\overline{X}}=\left[\frac{1}{U-1} \sum_{u=1}^U \left(\overline{X}_{u} - \overline{\overline{X}}\right)^2 \right]^{-1}
\qquad\mbox{and}\qquad
\tau_{\hat{\tau}}=\left[\frac{1}{U-1} \sum_{u=1}^U \left(\hat{\tau}_u - \overline{\hat{\tau}} \right)^2 \right]^{-1}.\\
$$
The covariate vector $(\mu_1, \ldots, \mu_U)$ was also summarised by its mean and precision, $\overline{\mu}$ and $\hat{\tau}_{\mu_u}$
$$
\overline{\mu} = \frac{1}{U}\sum_{u=1}^U \mu_u \quad \text{and} \quad
\hat{\tau}_{\mu_u}=\left[\frac{1}{U-1} \sum_{u=1}^U \left(\mu_u - \overline{\mu} \right)^2 \right]^{-1}.\\
$$
Sampling from the full conditional distribution of $\tau_x$ is achieved following the same procedure as for $\mu_u$, except that we use a gamma (rather than Gaussian) neural network model for the non-linear mean function.

The essential idea behind the ABC-PaSS method \shortcite{Kousathanas2016} is to use approximately conditionally sufficient  summary statistics within low-dimensional conditional Metropolis-Hastings updates.
To conduct a fair comparison with  approximate Gibbs sampling, to update $\tau_x$ and $\mu_u$, at each iteration we draw proposals from their known (in this case) full conditional distributions. 
This favourably gives ABC-PaSS the best possible proposal distribution,
and so allows the comparison between algorithms to focus on the form of the update mechanism. 
The summary statistics used for each parameter update are the same as for the approximate Gibbs samplers ($\bs{S}_\tau$ and $\bs{S}_u$ for $\tau_x$ and $\mu_u$ respectively). 
Generating $\bs{S}_u$ only requires simulating data from group $u$.
The updates for $\mu$ and $\tau_\mu$ are performed using Gibbs updates, as before.
We consider a single `iteration' of the ABC-PaSS algorithm to update each model parameter in turn.

We generate $L=10$ observations from $U=10$ groups with $\mu=0$, $\tau_{\mu}=\tau_x=1$ and $\alpha_\mu=\nu_\mu=\alpha_x=\nu_x=1$. We simulate $M=10,000$ iterations from each sampler.
For the approximate Gibbs samplers, we first generated $N=10,000$ synthetic datasets from the prior predictive distribution. For the global models we chose $K_h$ to be uniform, with $h$ determined to select the closest $5,000$ samples (in terms of Euclidean distance) to the observed \emph{symmetric} summary statistics. For the local model, for each localised regression model we kept the closest 10\% of the $5,000$ samples. 
For the kernels $K_h$ in the Metropolis-Hastings updates of the ABC-PaSS algorithm we set $h=0.5, 2$ for $\mu_u$ and $\tau_x$ respectively.
Each simulation was replicated a total of 500 times.

\begin{figure}[!h]
	\centering
	\subfloat[\footnotesize{Relative MSE}. \label{Acurracy}]{\includegraphics[width=5cm,height=7cm,angle=-90]{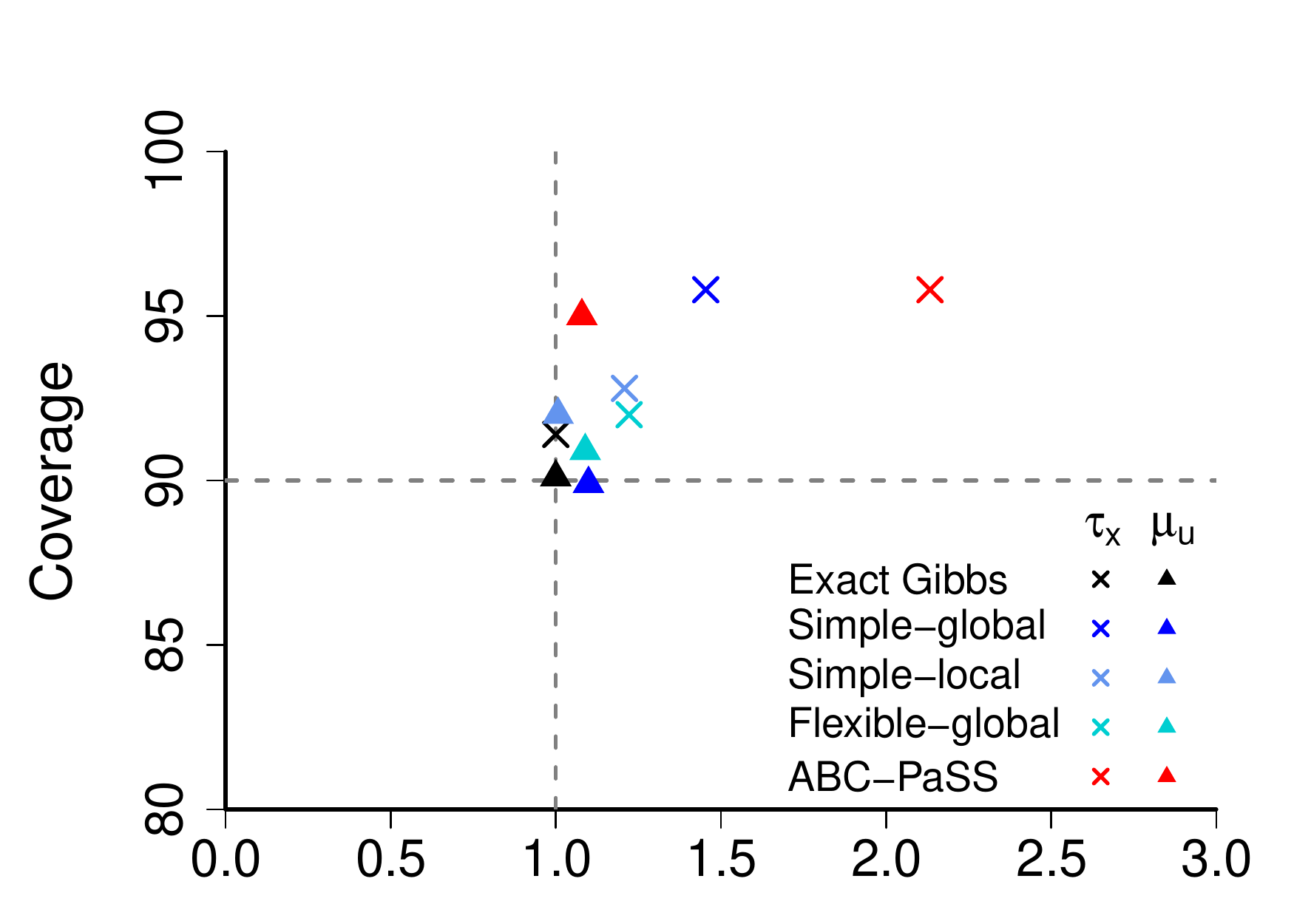}} 
	\subfloat[\footnotesize{Computational times}. \label{Times}]{\includegraphics[width=5cm,height=7cm,angle=-90]{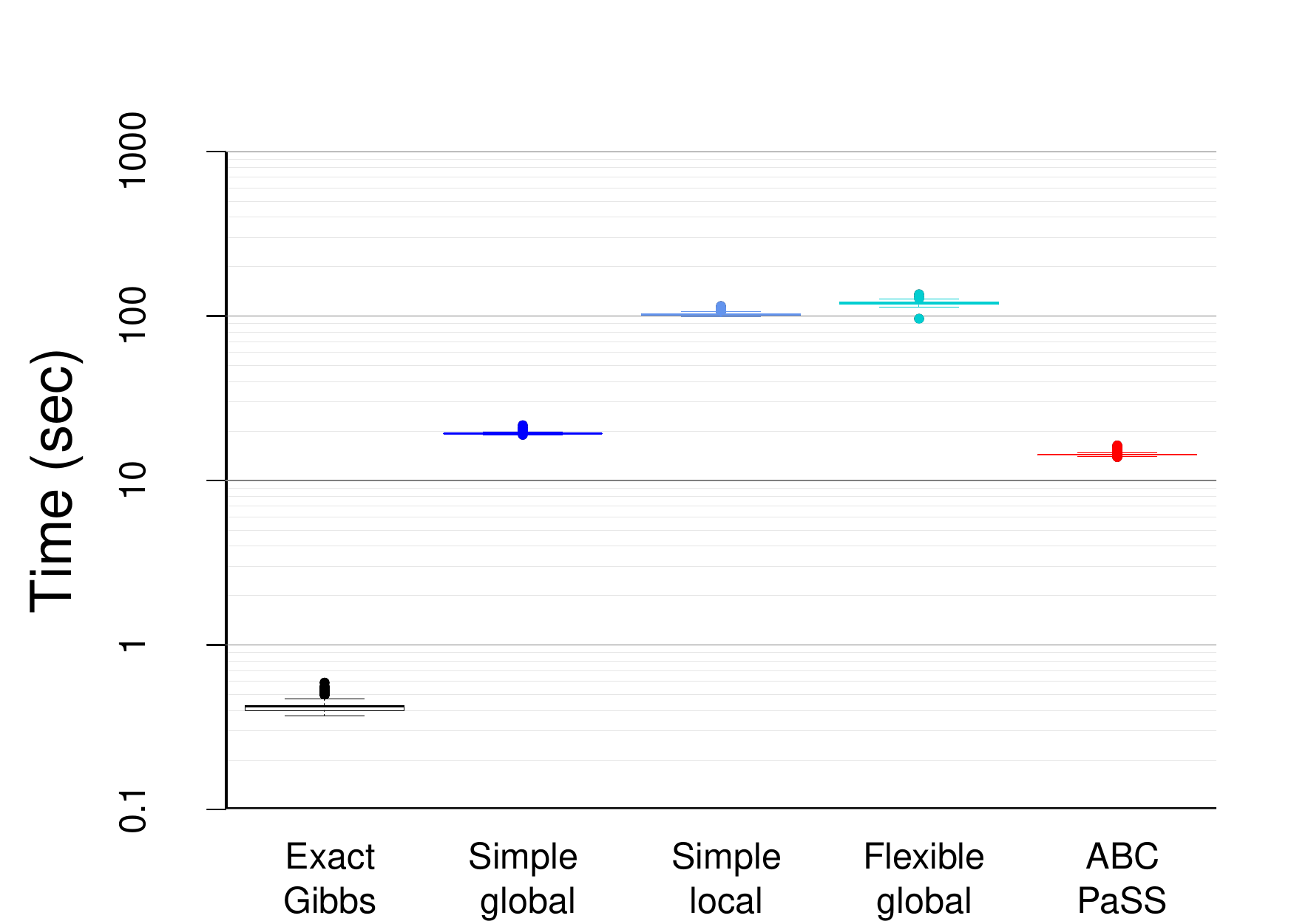}}
	\\
	\subfloat[\footnotesize{Posterior for $\mu_1$}. \label{mu_1}]{\includegraphics[width=5cm,height=7cm,angle=-90]{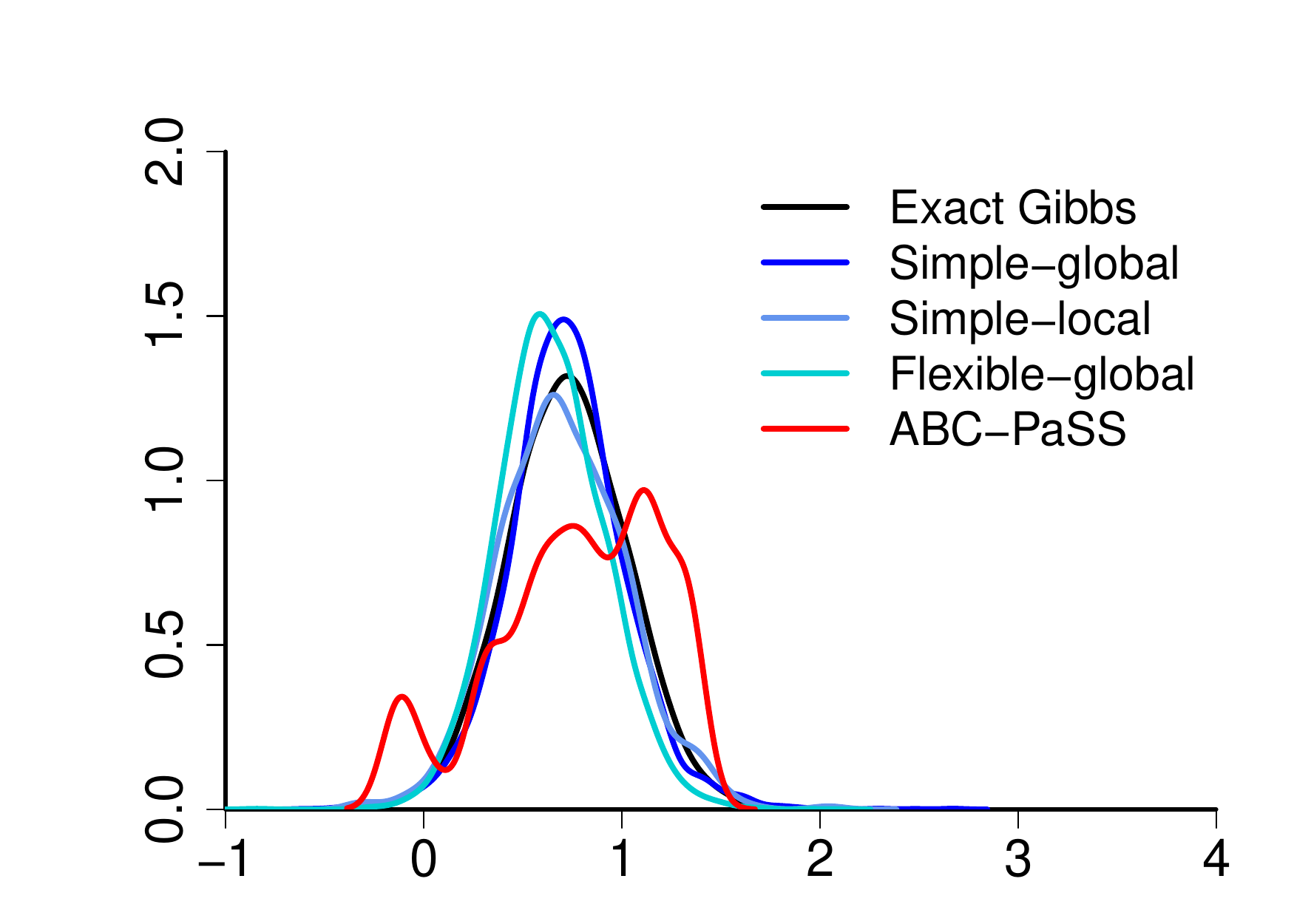}}
	\subfloat[\footnotesize{Posterior for $\tau_x$}. \label{tau_x}]{\includegraphics[width=5cm,height=7cm,angle=-90]{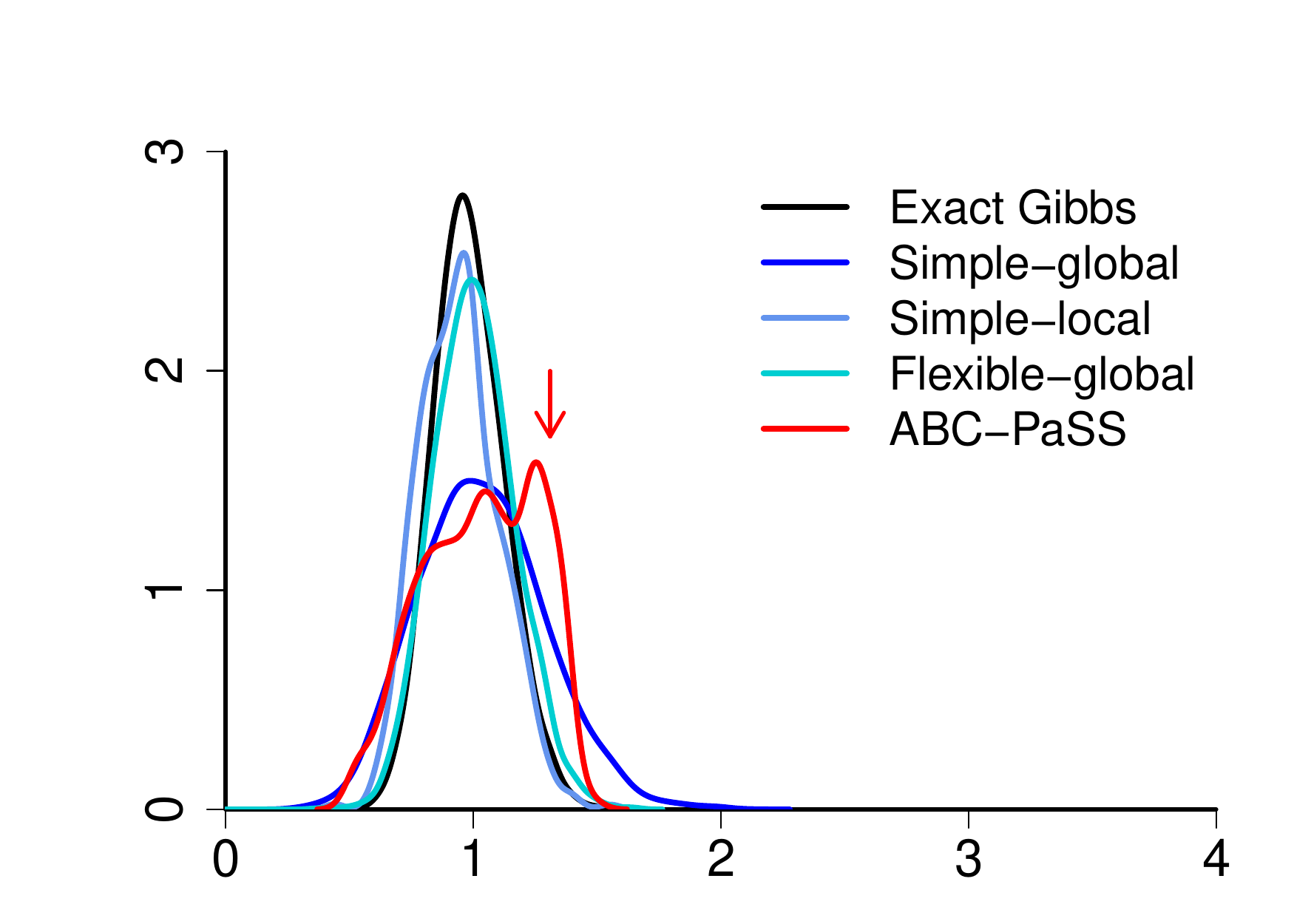}}
    \\
	\subfloat[\footnotesize{ABC-PaSS sampler}. \label{chain-ABC-PaSS}]{\includegraphics[width=5cm,height=7cm,angle=-90]{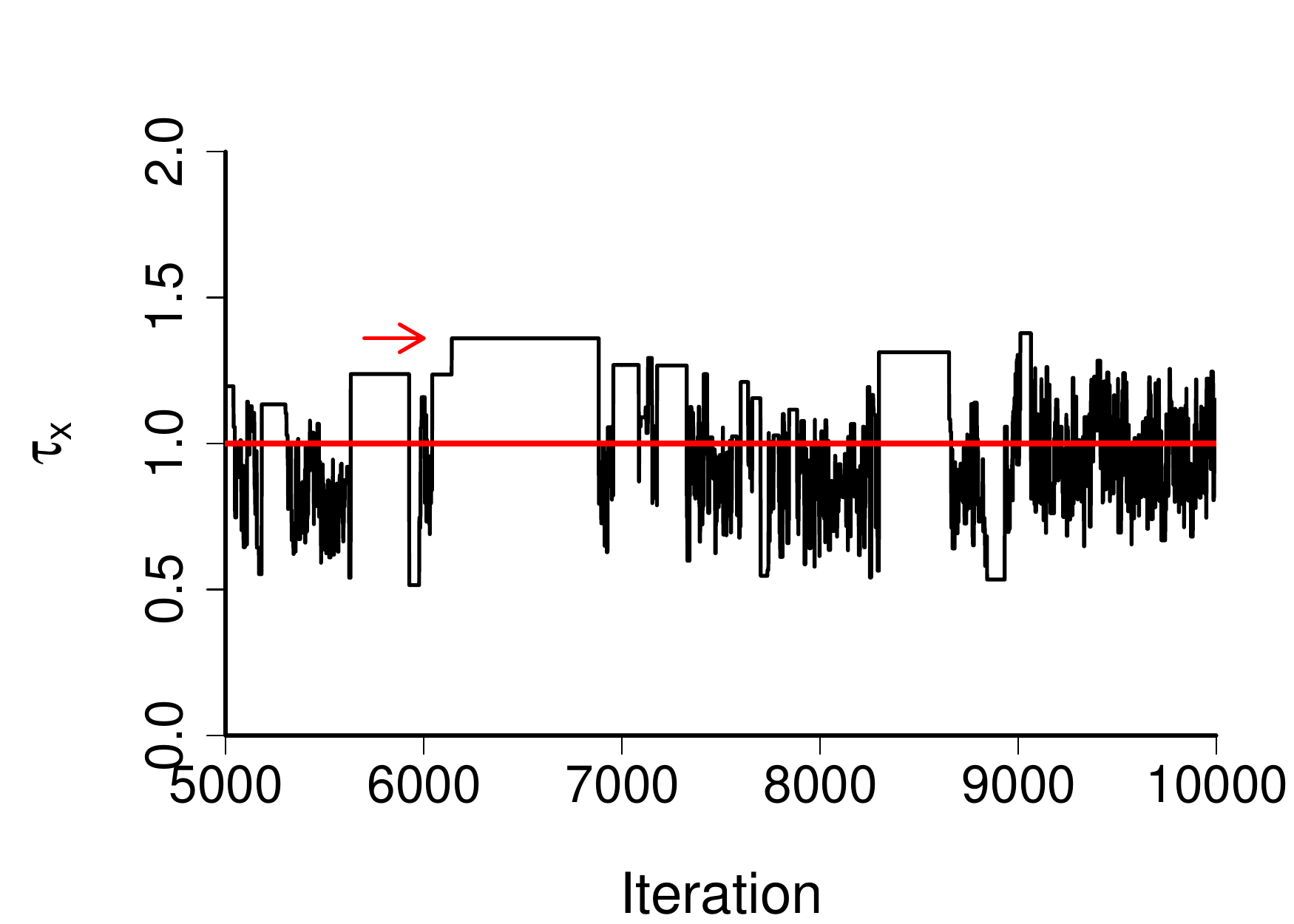}}
	\subfloat[\footnotesize{Flexible-global approx.~Gibbs sampler}. \label{chain-DL}]{\includegraphics[width=5cm,height=7cm,angle=-90]{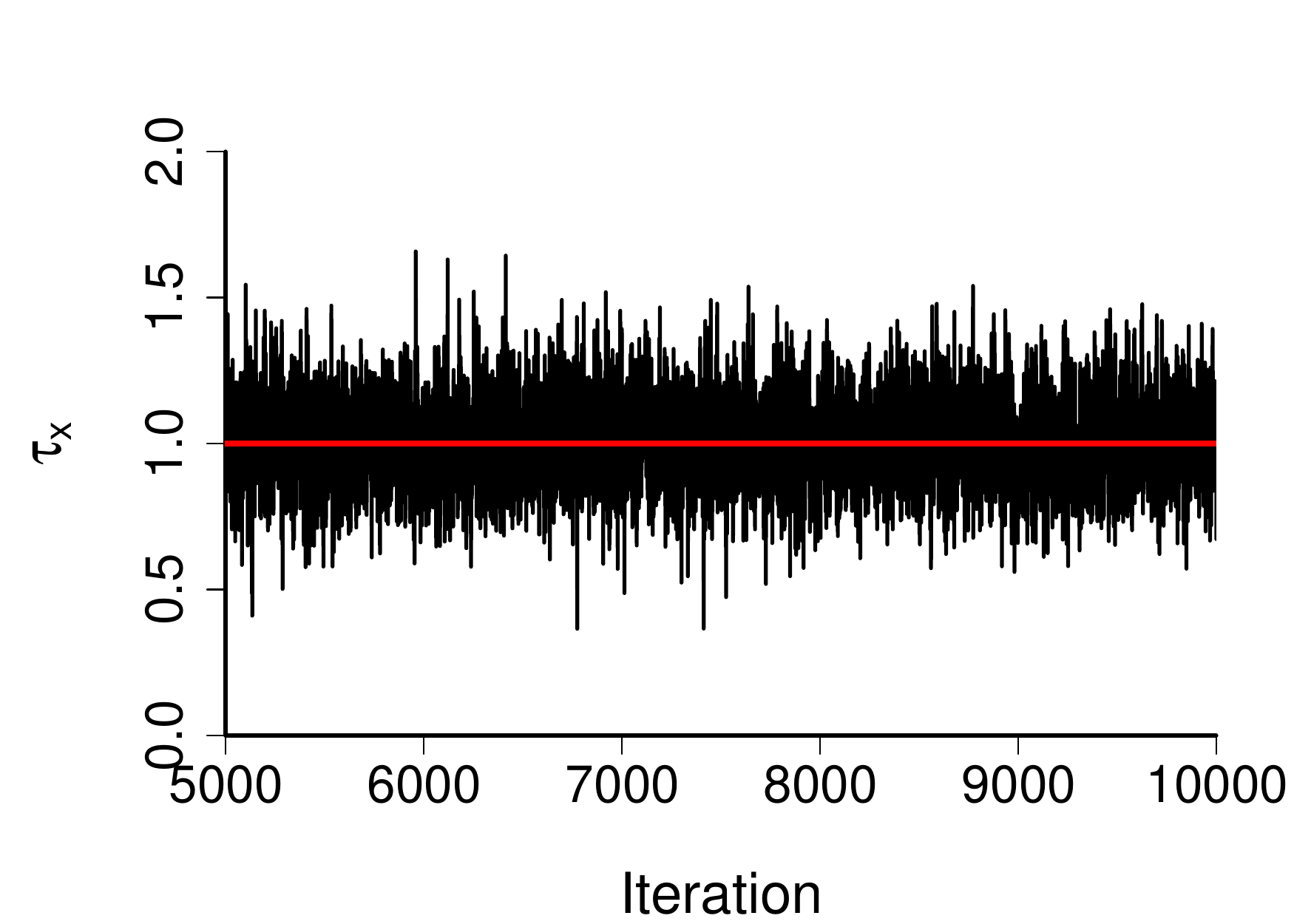}}
	\caption{\small Performance of the exact Gibbs sampler, approximate Gibbs samplers (shades of blue) and the modified ABC-PaSS sampler (red) for the simple hierarchical model. (a) Mean relative MSE and observed ($90\%$) coverage for $\tau_x$ and $\mu_u$ based on 500 replications.  
	(b) Log-scale boxplot of sampler implementation times.
	(c, d) Estimated marginal posteriors for $\mu_1$ and $\tau_x$ for a single simulation.
	(e, f) Typical Markov chain (only the last 5000 iterations, for clarity) sample paths for $\tau_x$ for the ABC-PaSS and the flexible-global approximate Gibbs sampler.
}
	\label{fig:simulation}
\end{figure}

Figure \ref{Acurracy} shows the relative (mean) MSE
and observed coverage 
$90\%$ credibility intervals for $\tau_x$ and $\mu_u$ with respect to the exact Gibbs sampler. ABC-PaSS performed significantly worse than the other samplers.
For the approximate Gibbs samplers
the simple global model performed well for $\mu_u$, but was clearly worse for $\tau_x$, when compared to the other model specifications which performed relatively well.

Panel \ref{Times} illustrates the time taken to run each sampler. The exact Gibbs sampler takes less than 1s to complete, while the simple global approach and ABC-PaSS take less than 20s on average. The remaining methods had comparable times ($\sim$100s). 
These times are broken down in Table \ref{tab:Times}.
The ABC-PaSS algorithm does not fit regression models and dataset generation is performed within the MCMC sampler. For this example, generating 10,000 synthetic datasets took only 6.58 seconds. The flexible-global approach required the fit of four Deep Learning regression models (modelling both mean and variance of $\tau_x$ and $\mu_u$),
each computationally expensive. Whereas the simple-local approach required 20,000 regression model fits, making each sampler iteration 3.5 times slower than the flexible-global strategy. 
This simulation suggests that in applications where synthetic sampling is an expensive operation \shortcite{rodrigues+ps17}, ABC-PaSS will be largely inefficient. In comparison, the approximate Gibbs samplers make more efficient and repeated use of each synthetic sample, within each sampler iteration.
In practice the optimal approach will be determined by balancing the cost of synthetic dataset simulation and the required number of MCMC samples.

\begin{table}[ht]
\centering
\begin{tabular}{|l||cc|cc||c|}
 \hline
 Method & \multicolumn{2}{c|}{Synthetic samples} & \multicolumn{2}{c||}{Regression fits} & MCMC\\ 
 &Time (s) & Number & Time (s) & Number & Sampler\\
 \hline 
 Exact Gibbs & 0  &0 & 0 &0 & \phantom{1}0.42 \\ 
 Simple-global & 6.58 &10,000\phantom{$^*$}  & \phantom{8}0.02 &2 & 12.75 \\ 
 Simple-local & 6.58 &10,000\phantom{$^*$}  & 0 & 20,000$^*$ & 95.77 \\ 
 Flexible-global & 6.58 &10,000\phantom{$^*$} & 85.85  &4 & 27.51 \\ 
 ABC-PaSS & 0 &20,000$^*$ & 0 & 0 & 14.38 \\ 
 \hline
\end{tabular}
\caption{\small Mean time (seconds) and number of synthetic dataset generations for each sampler, based on 500 replicate chains. Figures for synthetic samples and regression fits are for operations {\em before} the sampler is run, excluding those indexed by $*$, which are performed within the sampler.
}
\label{tab:Times}
\end{table}

Figures \ref{mu_1} and \ref{tau_x} show the estimated marginal posterior densities for $\mu_1$
and $\tau_x$.
All approximate Gibbs implementations reasonably estimate the true density (black line), but the performance of ABC-PaSS is clearly poor. For this algorithm, by setting the $K_h$ kernel scale parameter to $h=0.5,2$ for $\mu_u$ and $\tau_x$ we achieved Metropolis-Hastings acceptance rates of 20\% and 18\% respectively. Lowering $h$ could improve the accuracy of this algorithm, however the sampler acceptance rates would fall further, and already the chain is experiencing poor mixing (the  `sticking' phenomenon; \shortciteNP{sisson+ft07}) in the tail of the distribution (Figure \ref{chain-ABC-PaSS}). In contrast, mixing for the approximate Gibbs sampler is excellent (Figure \ref{chain-DL}).

Of the approximate Gibbs samplers, the simple-global approach performs least well for $\tau_x$ -- this is hardly surprising
given  the large differences between the exact conditional distributions and the simple regression models. However, localising the regressions (light blue line) at each stage of the Gibbs sampler produces a major improvement in the quality of the approximation. The same applies when the chosen regression models are flexible enough to accommodate non-linearities, interactions and heteroscedasticity (dark blue line).

  %%%%%%%%%%%%%%%%%%%%%%%%%%%%%%%
  %%%%%%%%%%%%%%%%%%%%%%%%%%%%%%%
  \section{A state space model of {\em Airbnb} data} 
  %%%%%%%%%%%%%%%%%%%%%%%%%%%%%%%
  %%%%%%%%%%%%%%%%%%%%%%%%%%%%%%%
 \label{application}
  
  We analyse a time series dataset containing Airbnb property rental prices in the city of Seattle, WA, USA in 2016. The dataset,
  available at \emph{kaggle.com},
  consists of 928,151 entries, each corresponding to an available listed space (property, room, etc) at a given date. 
  The price distribution of these data on each day is non-Gaussian 
  even after transformation.
  Hence we use the more flexible
$g$-and-$k$ distribution \cite{Haynes1998,Rayner2002}, which has an intractable density function, 
but a tractable quantile function 
  \[
   Q(q|A, B, g, k) = A + B \left[1 + c\frac{1-\exp\{-gz(q)\}}{1+\exp\{-gz(q)\}} \right] (1+z(q)^2)^k z(q), 
  \]
  for $B>0$ and $k>-0.5$ (with $c=0.8$), where $z(q)$ denotes the $q$-th quantile of the standard Gaussian distribution. 
As a simple 4-parameter univariate model  with an intractable density, this distribution has gained popularity in the ABC literature \cite{Drovandi2011,Fearnhead2012,peters+s06}. 
  Figure \ref{fig:observed} shows $L$-moments estimates of each 
  $g$-and-$k$ 
  parameter  \shortcite{Peters2016} for each day in the {\em Airbnb} dataset. Each parameter exhibits  a dynamic level with a weekly seasonal effect, and a sudden shift induced by the start and end of the extended summer season (1st April to 31st September), as well as additional stochastic variation potentially depending on other factors.
The series are also dependent with e.g.~a strong negative correlation between scale ($B$) and kurtosis ($k$).

        \begin{figure}[htp]
	\centering
	\subfloat[\footnotesize{Location}\label{fig:A}]{\includegraphics[width=4.5cm,height=4.5cm,angle=-90]{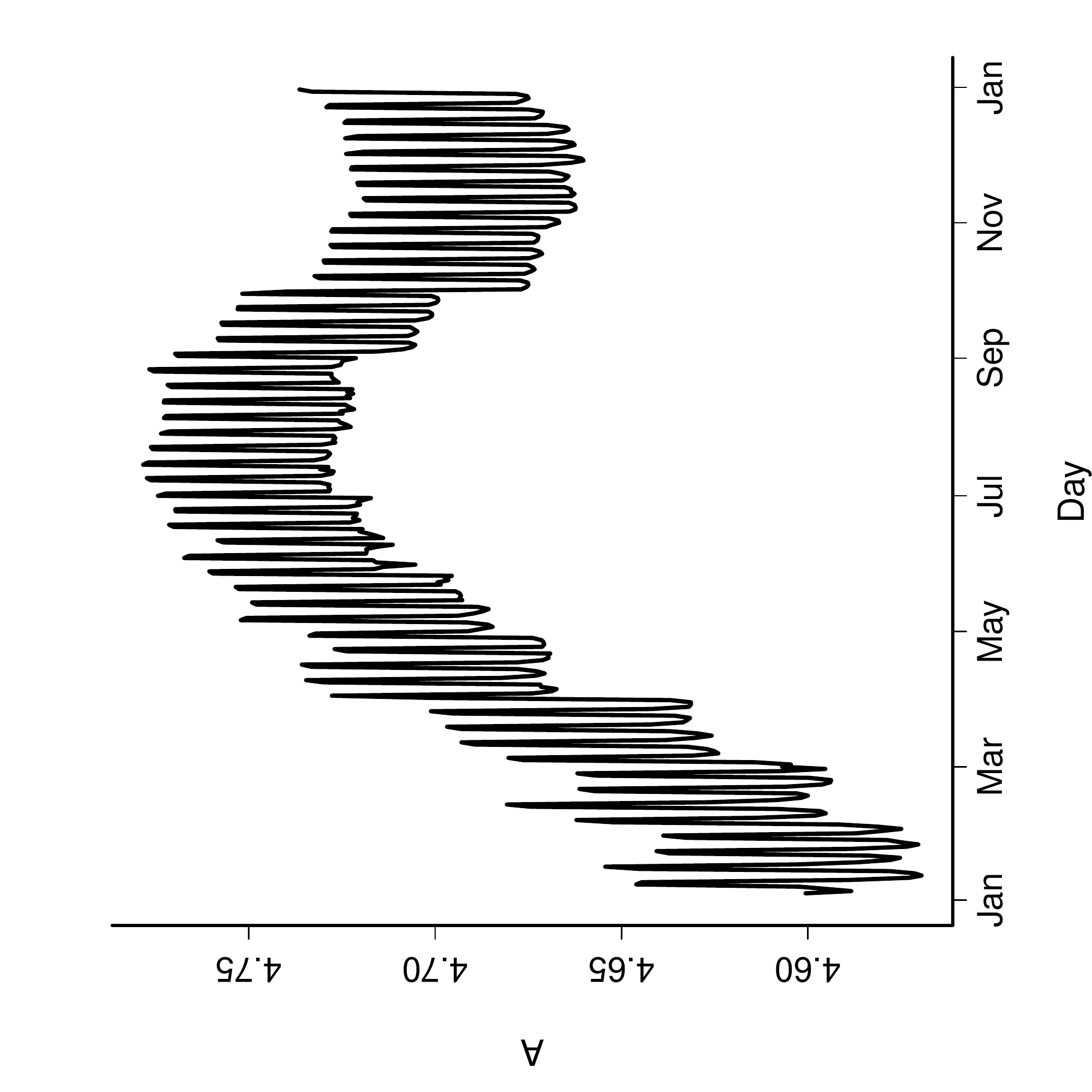}} 
	\subfloat[\footnotesize{Scale}\label{fig:B}]{\includegraphics[width=4.5cm,height=4.5cm,angle=-90]{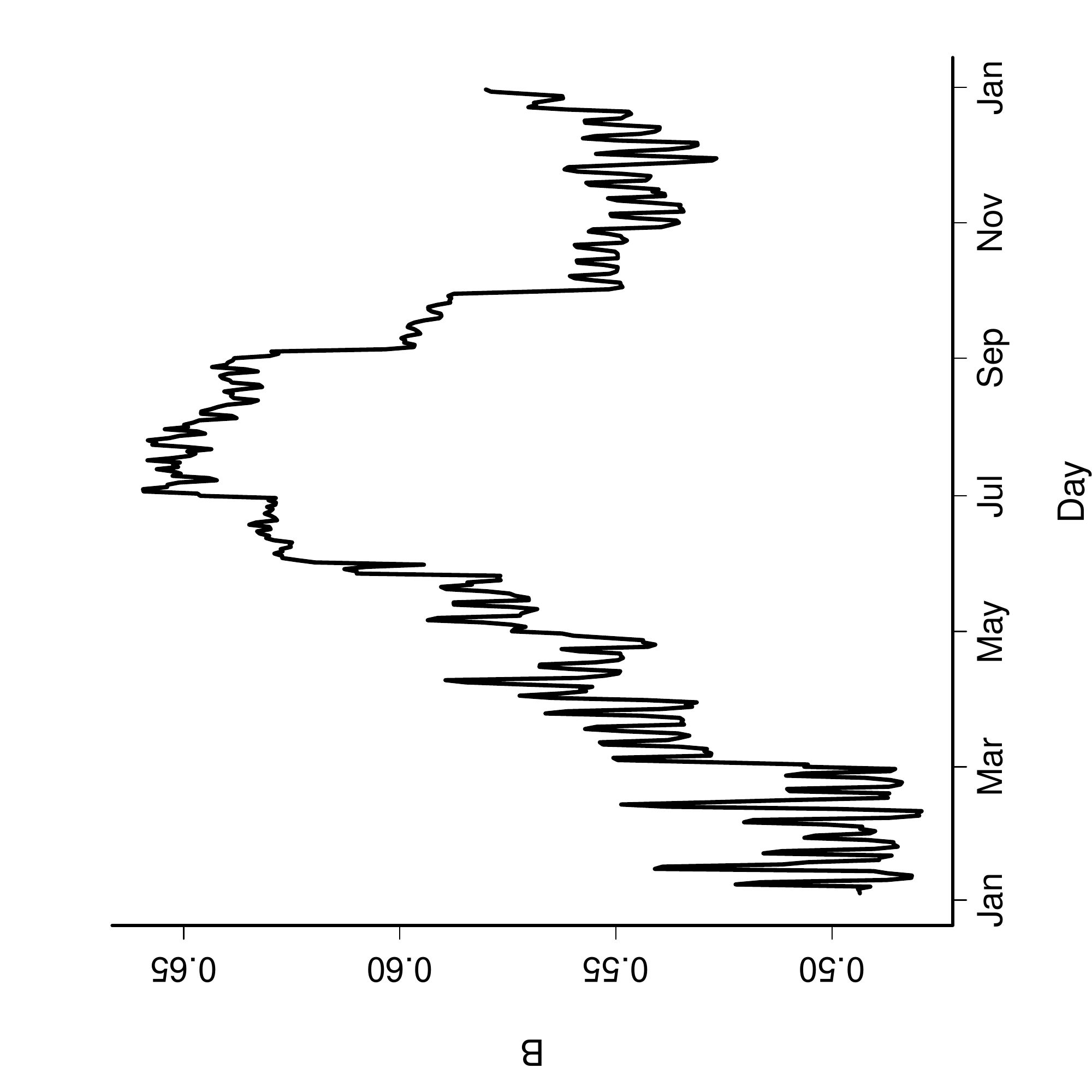}}
	\subfloat[\footnotesize{Skewness}\label{fig:g}]{\includegraphics[width=4.5cm,height=4.5cm,angle=-90]{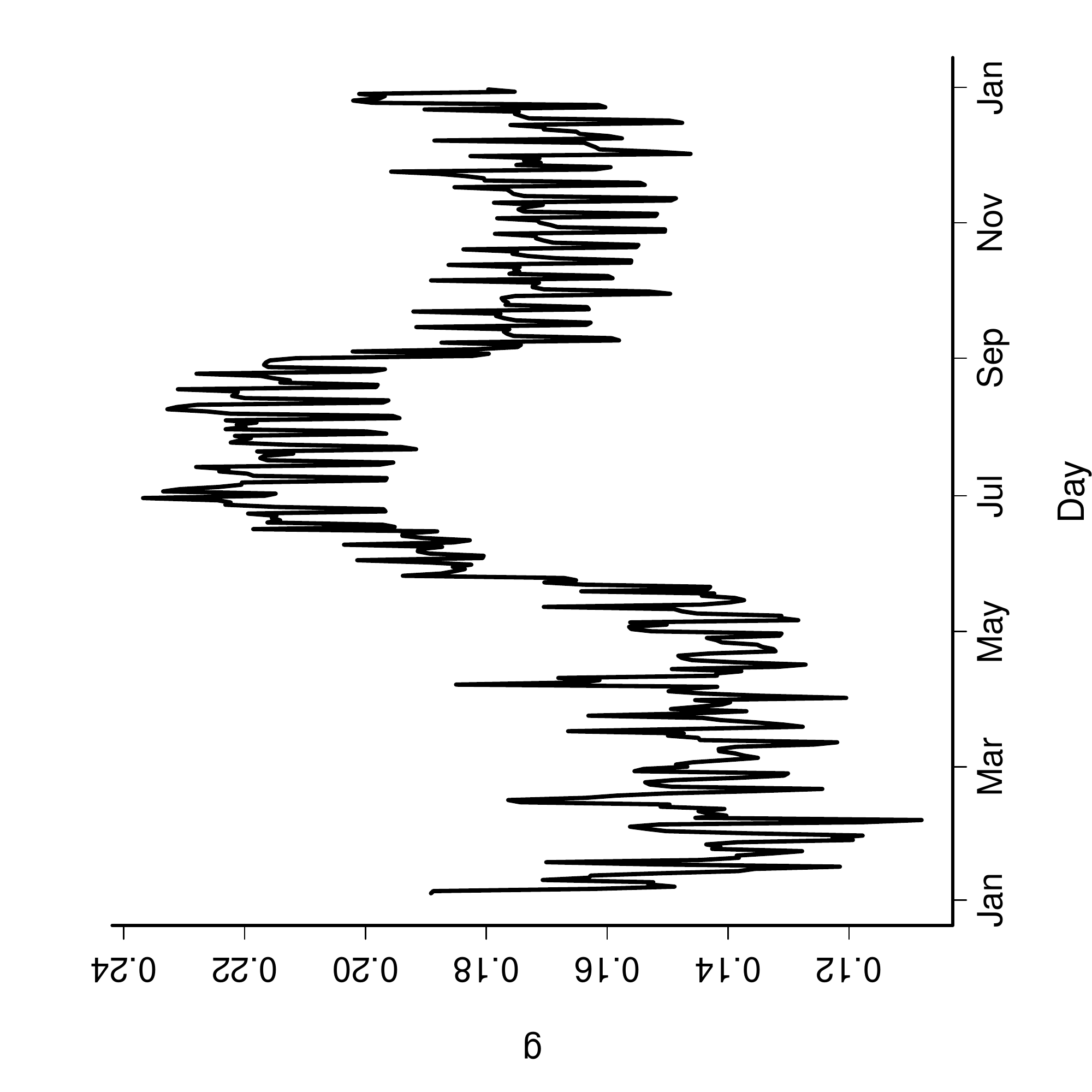}}
	\subfloat[\footnotesize{Kurtosis}\label{fig:k}]{\includegraphics[width=4.5cm,height=4.5cm,angle=-90]{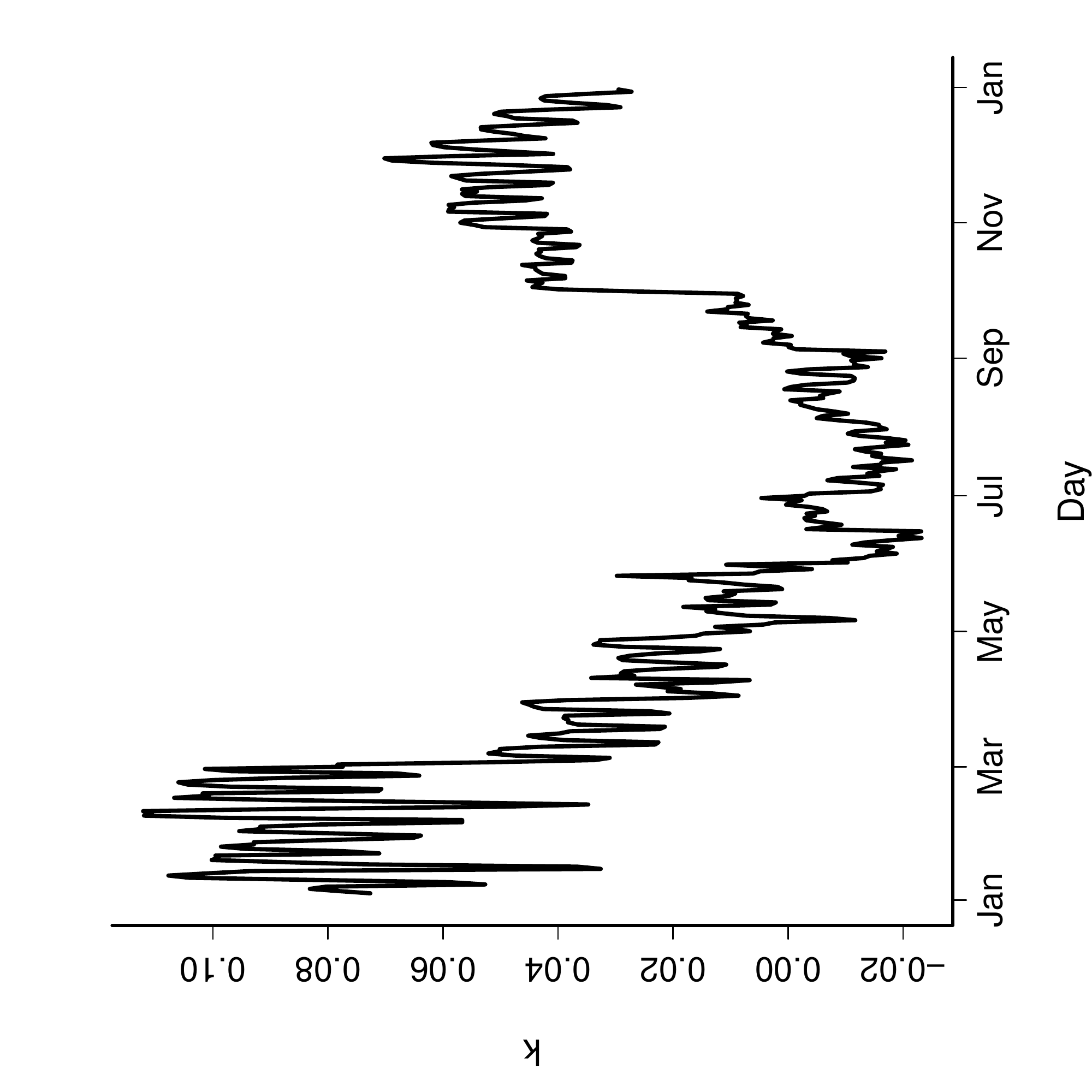}}
	\caption{\small L-moments $g$-and-$k$ distribution parameter estimates ($A, B, g, k$) for each day in the {\em Airbnb} dataset (Peters et al. 2016).
	}
	\label{fig:observed}
      \end{figure}

We construct the following intractable non-linear state space model:
  \begin{subequations}
  \begin{flalign}
   \text{Observation distribution:} && \bfy_t & \sim g\mbox{-and-}k(\bfbeta_t),\quad t=1, \ldots, T \\
   \text{Link function:} && h(\bfbeta_t) & = \bflambda_t = \bfF^\top_t \bftheta_t \\
   \text{System equation:} && (\bftheta_t|\bftheta_{t-1}) & = \bfG_t \bftheta_{t-1} + \bfw_t, \quad        \bfw_t \sim N(\mathbf{0}, \bfW_t) \\ 
   \text{Prior distribution:} && \bftheta_0 & \sim N(\bfm_0, \bfC_0), 
  \end{flalign}
  \label{eq.model}
  \end{subequations}
\noindent  where $\bfy_t$ denotes the vector of (log) prices observed at time $t$, $\bfF_t$ is a known $p \times 4$ design matrix that maps the state vector $\bftheta_t$ to the linear predictor $\bflambda_t=(\lambda_{1,t},\ldots,\lambda_{4,t})^\top$, $\bfG_t$ is a known $p \times p$ evolution matrix that dictates the system's dynamics, $\bfW_t$ is a possibly unknown covariance matrix, and $\bfbeta_t = (\lambda_{1, t}, \exp(\lambda_{2, t}), \lambda_{3, t}, \exp(\lambda_{4, t})-0.5)^\top=(A, B, g, k)^\top_t$ represents the $g$-and-$k$ distribution parameters. The link function $h(\cdot)$ ensures that $\bfbeta_t$ respects the constraints imposed by the observation distribution. We assume that given $\bftheta_t$, the observations $\bfy_t$ are independent and identically distributed. The sequence of errors $\bfw_t$ are also assumed to be independent. 
Specification of $\bfF_t$ and $\bfG_t$ is provided in Appendix A.1.
For this analysis we set $\bfm_0 = \bf 0$ and $\bfC_0 = 10^{7} \bf I$, where $\bf 0$ is a vector of zeros and $\bf I$ is the identity matrix,
%. 
 and $\bfW_t = \bfW = \text{diag}(1/\tau_1, \ldots, 1/\tau_p)$, with $\tau_i \sim \text{Gamma}(\alpha=10^{-10}, \nu=10^{-10})$, for $i=1, \ldots, p$.

State space models provide a flexible and well-structured framework to probabilistically describe an extensive array of applied problems
\shortcite{West1997,Petris2010}.
\shortciteN{West1985} introduced dynamic generalised linear models, which relaxed the linearity and Gaussian assumptions, allowing the observations to follow other members of the exponential family. Other works have focused on specific observation distributions, such as the Beta \shortcite{da-Silva2011} and the Dirichlet \shortcite{da-Silva2013}. 
  Computational hurdles have limited the use of intractable dynamic models such as the one considered here, but  increasing efforts to tackle this issue are being made \shortcite{Jasra2012,Dean2014,Martin2014,Calvet2012,Yildirim2013,Picchini2016,Martin2016}. Our approach extends the method given by \shortciteN{Peters2016}.

  Writing $\bfTheta=(\bftheta_0, \ldots, \bftheta_T)$, the joint distribution factorises as
  \begin{align} 
    \label{eq.joint}   p(\bfTheta, \bfW, \bfy_1, \ldots, \bfy_T) = p(\bftheta_0) p({\bfW}) \prod_{t=1}^T 
    \left[ p(\bftheta_t|\bftheta_{t-1}, \bfW) p(\bfy_t|\bflambda_t) \right].
  \end{align}
The data $\bfy_t$ only depend on the system state through $\bflambda_t$, so the full conditional distribution for $\bftheta_t$ can be conveniently factorised as
  \begin{align*}
  p(\bftheta_t|\cdot) & = 
  p(\bftheta_t | \bftheta_{t-1}, \bftheta_{t+1}, \bfW, \bflambda_t) p(\bflambda_t | \bftheta_{t-1}, \bftheta_{t+1}, \bfW, \bfy_t).
  \end{align*}
 One can sample from this distribution in two stages: $\bflambda_t^* \sim p(\bflambda_t | \cdot)$ and then $\bftheta_t^* \sim p(\bftheta_t | \bflambda_t^*, \cdot)$. 
 All full conditional distributions are tractable (see Appendix A.2) apart from $p(\bflambda_t | \cdot)$.

  To approximate the linear predictor's conditional distribution, $p(\bflambda_t | \bftheta_{t-1}, \bftheta_{t+1}, \bfW, \bfy_t)$, 
  we reduce the dimension of the conditioning set  by replacing the observed data $\bfy_t$ by the summary statistic $\bfs_t=g(\hat{\bfbeta_t})$, where $\hat{\bfbeta_t}$ is the L-moments estimator of $\bfbeta_t$ given $\bfy_t$ and $g(\cdot)$ is the link function defined above.
  While not fully sufficient, these statistics are highly informative and nearly unbiased for all sample sizes and parameters  \shortcite{Peters2016}.

  It is useful to recognise that $p(\bflambda_t | \bftheta_{t-1}, \bftheta_{t+1}, \bfW, \bfs_t) = p(\bflambda_t | \bfphi_t, \bfs_t)$, where $\bfphi_t = (\bff_t, \bfq_t, n_t)$, and where $\bff_t = \bfF^\top_t \bfa_t,$ 
      $\bfq_t = F^\top_t \bfR_t F_t,$ and $n_t$ is the sample size at time $t$. As this structure is valid throughout the evolution period, the time label can be effectively dropped, which reduces the problem to approximating the distribution of a 4-dimensional vector, $\bflambda$, conditional on 13 variables ($\bfq_t$ is a diagonal matrix).  
  Without loss of generality, we write
  \begin{align}
   \label{eq.lambda.dist}  (\bflambda | \bfphi, \bfs) = \bfmu_\lambda + \bfSigma_\lambda^{1/2} \bfepsilon_\lambda,
  \end{align}
  where $\bfmu_\lambda$ and $\bfSigma_\lambda^{1/2}$, as functions of $\bfphi$ and $\bfs$, respectively denote the mean and the (Cholesky) square root of the covariance of $(\bflambda | \bfphi, \bfs)$. $\bfepsilon_\lambda$ follows an unknown standardised distribution (that may also depend on $\bfphi$ and $\bfs$). Even without knowledge of the distribution of $\bfepsilon_\lambda$, given the moments of the joint vector,  
   {\small\begin{align*}
       \begin{pmatrix}
         \left.
         \begin{array}{c}
         \bflambda \\
         \bfs
         \end{array}
         \right|
       & \hspace{-.2cm} \bfphi
    \end{pmatrix}
    \sim 
    \begin{bmatrix}
      \begin{pmatrix}
        \bff \\
        \bff
      \end{pmatrix},
      & \bfOmega_{\bfphi} = 
      \begin{pmatrix}
        \bfOmega_{11} & \bfOmega_{12}  \\
        \bfOmega_{21} & \bfOmega_{22} 
      \end{pmatrix}
    \end{bmatrix}, 
    \end{align*}}
Linear Bayes (\citeNP{Hartigan1969,Goldstein1976}; and \shortciteNP{Nott2012} in an ABC context) can be employed to give the estimators
  \begin{equation}
    \hat{\bfmu}_\lambda  = \bff + \bfOmega_{12} \bfOmega_{22}^{-1} (\bfs - \bff) %\\
    \quad\mbox{and}\quad
   \hat{\bfSigma}_\lambda  = \bfOmega_{11} - \bfOmega_{12} \bfOmega_{22}^{-1} \bfOmega_{21}.
  \label{eq.moments}
\end{equation}
  To draw an approximate sample from $p(\bflambda_t | \bftheta_{t-1}, \bftheta_{t+1}, \bfW, \bfy_t)$ within the Gibbs sampler we a)
estimate the covariance matrix $\bfOmega_{\bfphi}$, b) compute the conditional moments in \eqref{eq.moments}, c) draw an approximate sample for $\bfepsilon_\lambda$, and d) plug-in the obtained values into \eqref{eq.lambda.dist}. 

To build the regression models we generate $N=5000$ samples of $\bfphi$ uniformly on a hypercube that roughly covers the region that might be visited during the Gibbs run: 
  the 
   means $\bff$ have the same range as observed in $\{\bfs_{obs,t}\}$, the diagonal elements of $\bfq$ are in the interval $(0, 10^{-5})$, and $n$ spans the observed sample sizes. See e.g.~\shortciteN{Fearnhead2012}, \shortciteN{fan+ns13} for other strategies.
For each sample $\bfphi^{(i)}=(\bff^{(i)}, \bfq^{(i)}, n^{(i)})$, $i=1, \ldots, N$, we draw $(\bflambda, \bfs)^{(i)} \sim (\bflambda, \bfs | \bfphi^{(i)}) = p(\bfs | \bflambda, n^{(i)}) p(\bflambda | \bff^{(i)}, \bfq^{(i)})$. 
Recall that $(\bflambda,\bfs)$ only depends on $t$ through $\bfphi$, so only a small single days' data needs to be generated.

For each step $m=1,\ldots,M$ in the approximate Gibbs sampler and for each $t=1,\ldots,T$, conditional on the current  value of $\bfphi^{*}_t$
we estimate 
   {\small 
   \begin{align*}
    \bfOmega_{\bfphi^{*}_{t}} = \text{V}
     \begin{pmatrix}
         \left.
         \begin{array}{c}
         \bflambda \\
         \bfs
         \end{array}
         \right|
       & \hspace{-.2cm} \bfphi^{*}_{t}
    \end{pmatrix}
    \approx \int \text{V}
     \begin{pmatrix}
         \left.
         \begin{array}{c}
         \bflambda - \bff \\
         \bfs - \bff
         \end{array}
         \right|
       & \hspace{-.2cm} \bfq, n
      \end{pmatrix}
      K_h(\|\bfphi-\bfphi^{*}_{t}\|) p(\bfphi) d\bfphi 
    \end{align*}
    }
by computing the kernel-weighted sample covariance matrix over the centered samples $(\bflambda^{(i)} - \bff^{(i)}, \bfs^{(i)} - \bff^{(i)})$, $i=1, \ldots, N$. 
We  used the Epanechnikov kernel $K_h$, with bandwidth chosen such that the closest $2000$ samples had non-zero weight.

For each $\hat{\bfOmega}_{\bfphi^{*}_t}$ we then compute $\hat{\bfmu}_{\lambda_t}^{*}$ and $\hat{\bfSigma}_{\lambda_t}^{*}$ from \eqref{eq.moments}. The empirical residuals are then given by $\bfepsilon_{\lambda}^{i, *} = (\hat{\bfSigma}_{\lambda_t}^{*})^{-1/2} (\bflambda^{(i)} - \hat{\bfmu}_{\lambda_t}^{*})$, $i=1,\ldots,N$. Finally, an approximate sample from the full conditional distribution 
$p(\bflambda_t | \bfphi^{*}_{t}, \bfs_t)$ is obtained by
\[
    \bflambda_t^{**} = \hat{\bfmu}_{\lambda_t}^{*} + (\hat{\bfSigma}_{\lambda_t}^{*})^{1/2} \bfepsilon_\lambda^{k, *}  \sim \int p(\bflambda_t | \bfphi, \bfs_t) K_h(\|\bfphi-\bfphi^{*}_t\|) p(\bfphi) d\bfphi,
\]
where the index $k$ is drawn from $(1, \ldots, N)$ with probability $\propto K_h(\|\bfphi^{(k)}-\bfphi^{*}_t\|)$.

   Figure \ref{fig:results} shows some of the estimated model components of $A_t$.
    In Figure \ref{fig:Levels},
    the deseasonalised posterior estimates (original scale) are plotted over the L-moment estimates, revealing the overall shape of the (location of the) price changes over the course of the year, with higher prices in the summer months.
   There is a clearly noticeable step change in prices for the duration of the high season.
   The season effect parameters
   are estimated to be effectively constant throughout  the high season, with $\hat{\bftheta}_{9, t}^{[1]} \approx 0.024$ for all such $t$. That is, prices are expected to uniformly increase by about $\exp(0.024)-1=2.4\%$ during the high season.
   
   \begin{figure}[H]
	\centering
	\subfloat[\footnotesize{Deseasonalised estimates}\label{fig:Levels}]{\includegraphics[width=5.5cm,height=5.5cm,angle=-90]{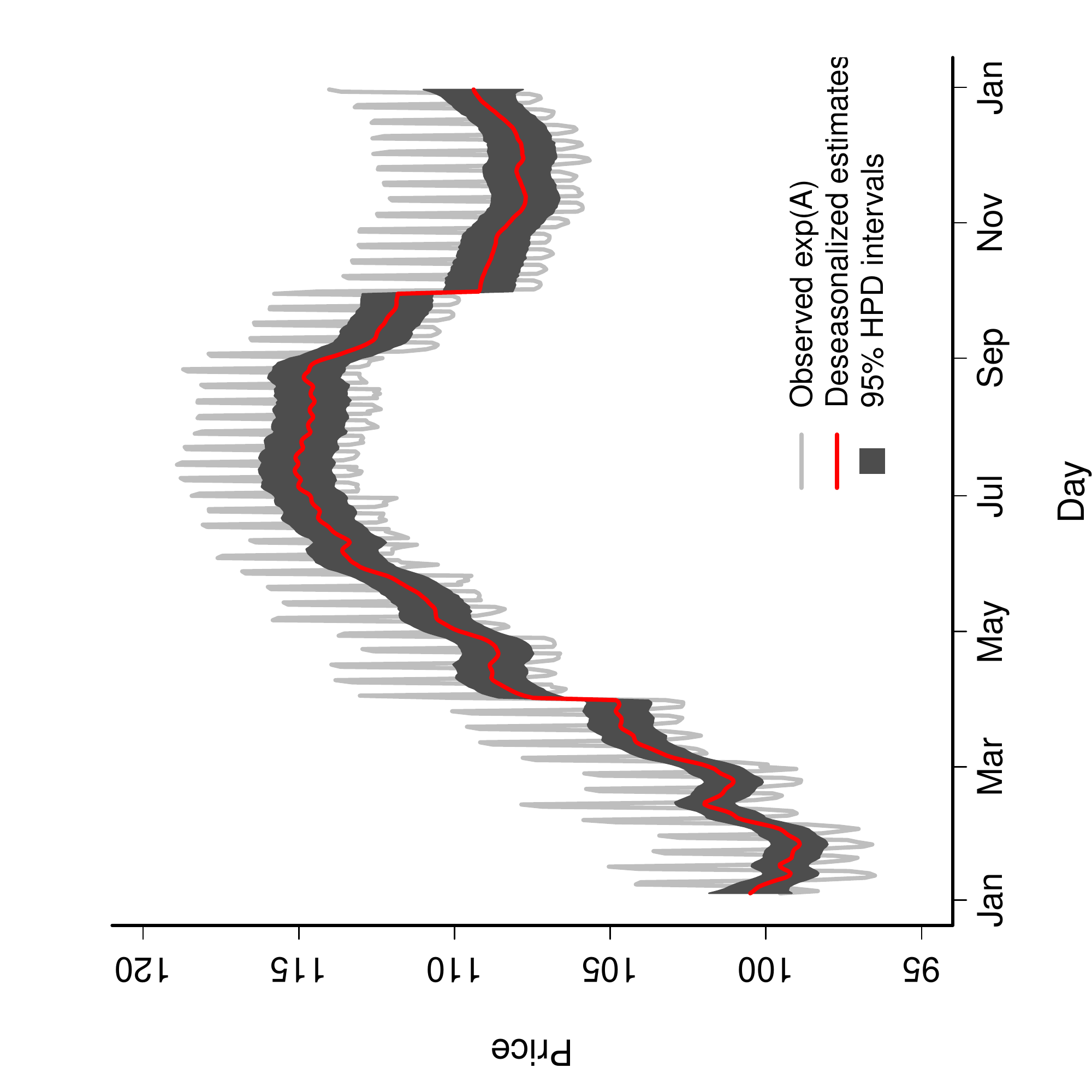}} 
	\subfloat[\footnotesize{Posterior estimates for $\exp(A_t)$}\label{fig:means}]{\includegraphics[width=5.5cm,height=5.5cm,angle=-90]{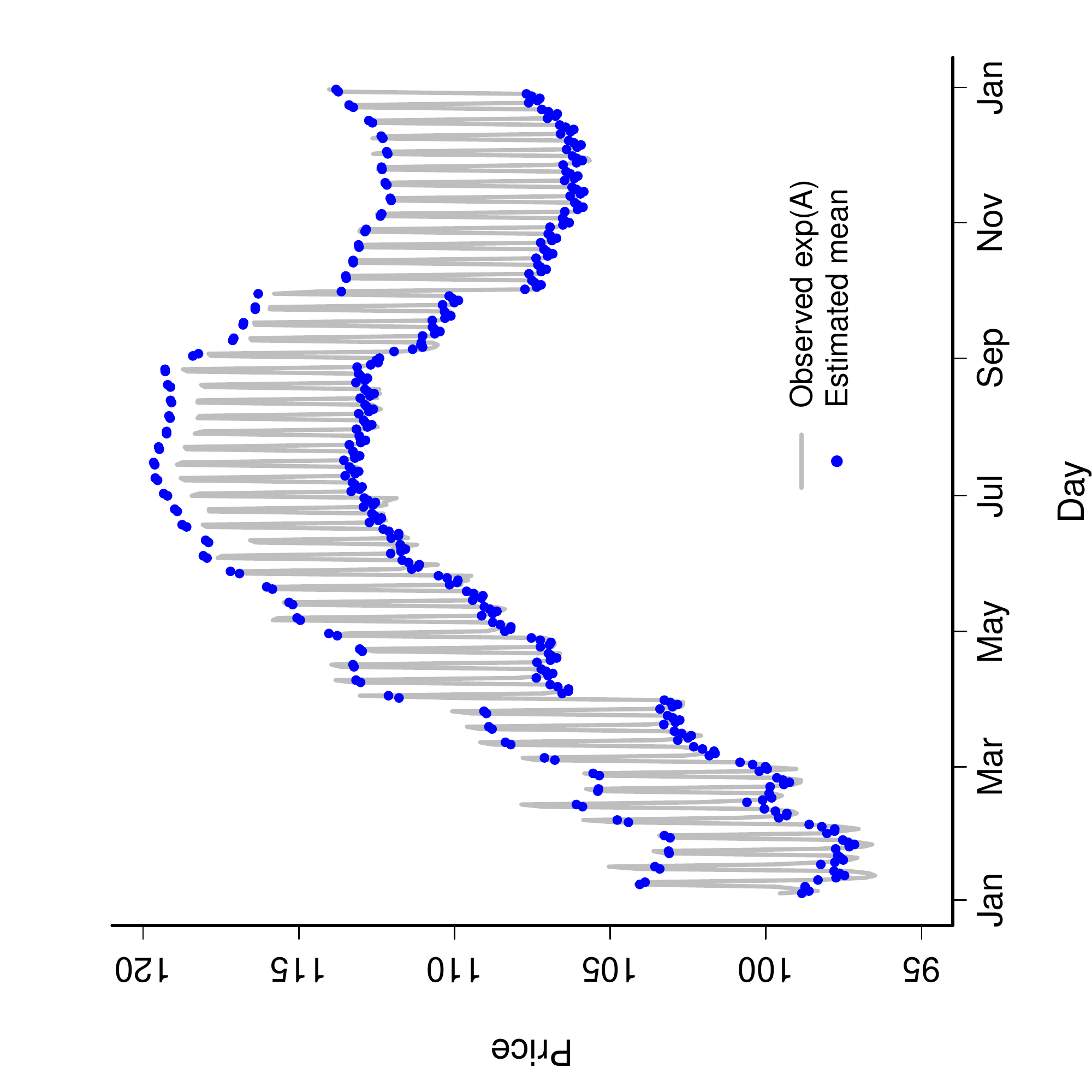}}
	\subfloat[\footnotesize{Estimated seasonal effect}\label{fig:Seasonal_effect}]{\includegraphics[width=5.5cm,height=5.5cm,angle=-90]{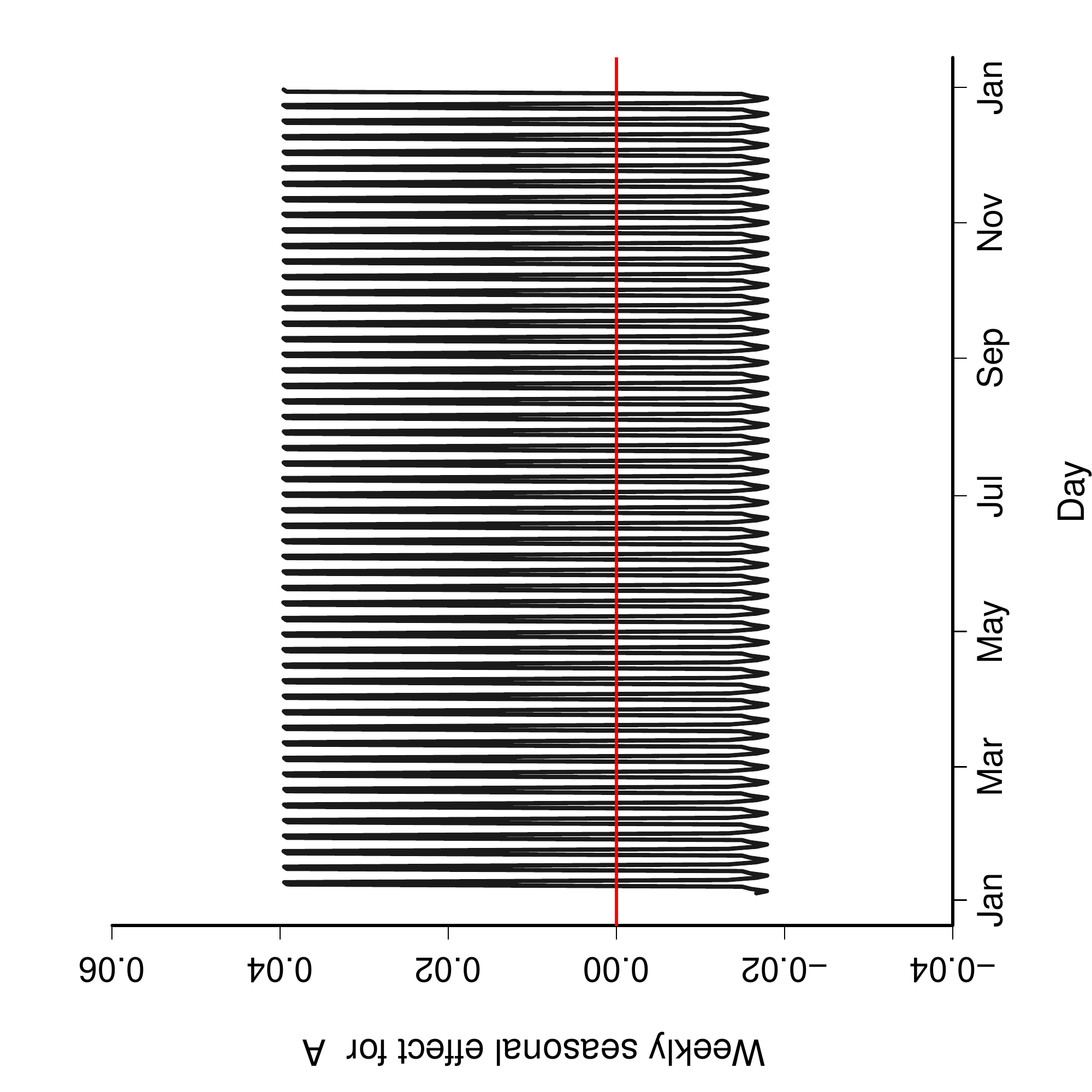}}
	\\
	\subfloat[\footnotesize{Residual plot}\label{fig:Residuals}]{\includegraphics[width=5.5cm,height=5.5cm,angle=-90]{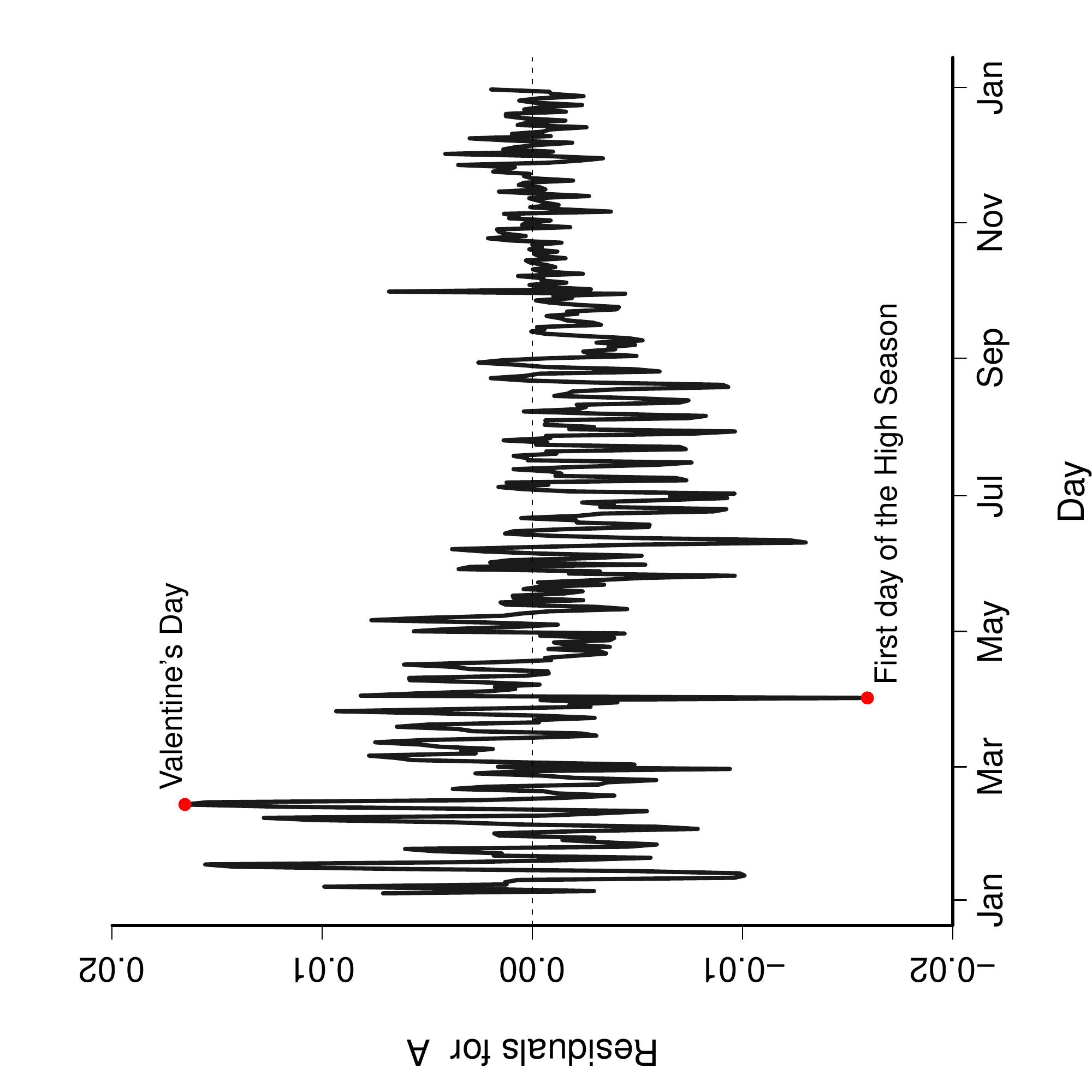}}
	\subfloat[\footnotesize{Parameter $A$ at time $t=1$.}\label{fig:Beta_trace}]{\includegraphics[width=5.5cm,height=5.5cm,angle=-90]{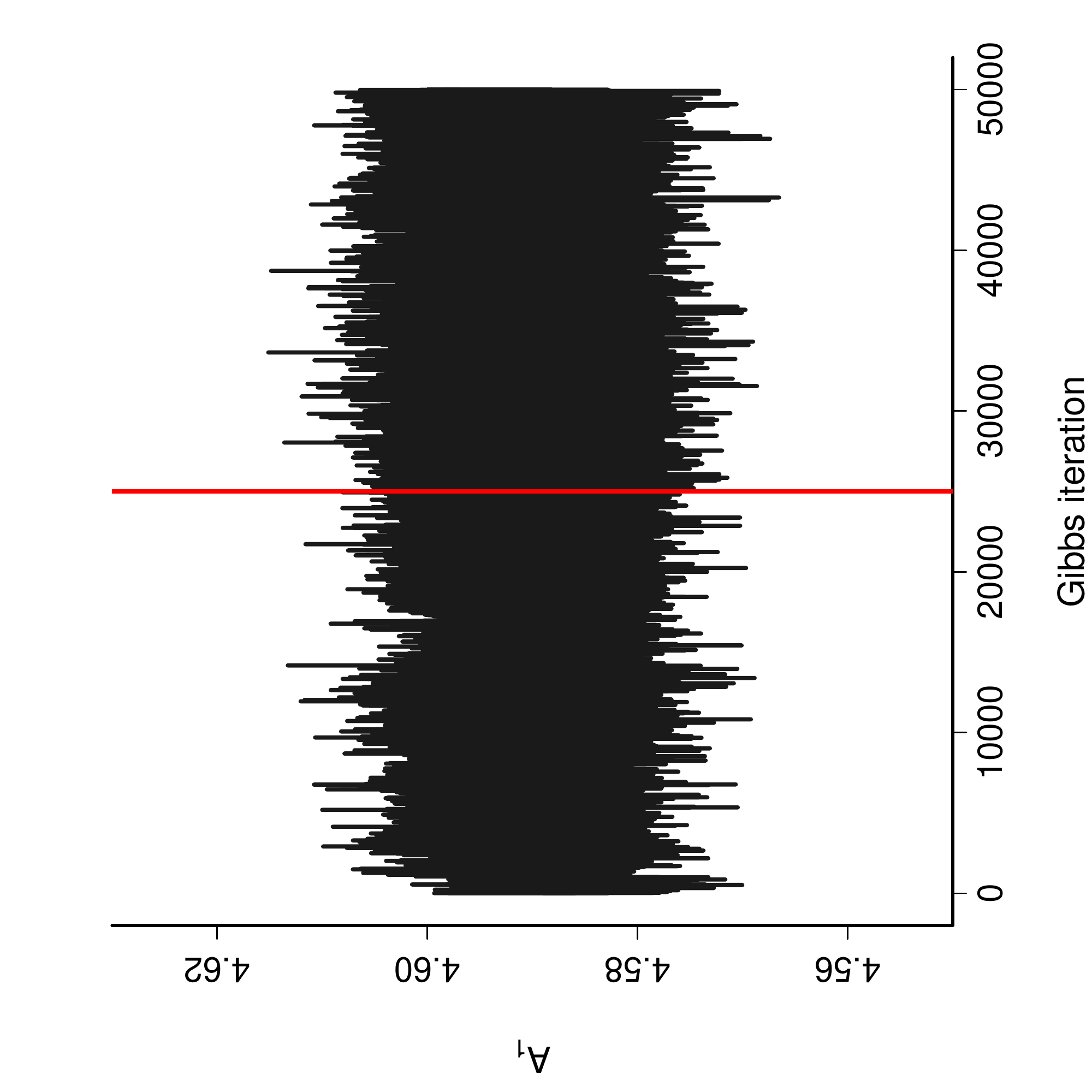}} 
	\subfloat[\footnotesize{Average season-effect}\label{fig:Summer_trace}]{\includegraphics[width=5.5cm,height=5.5cm,angle=-90]{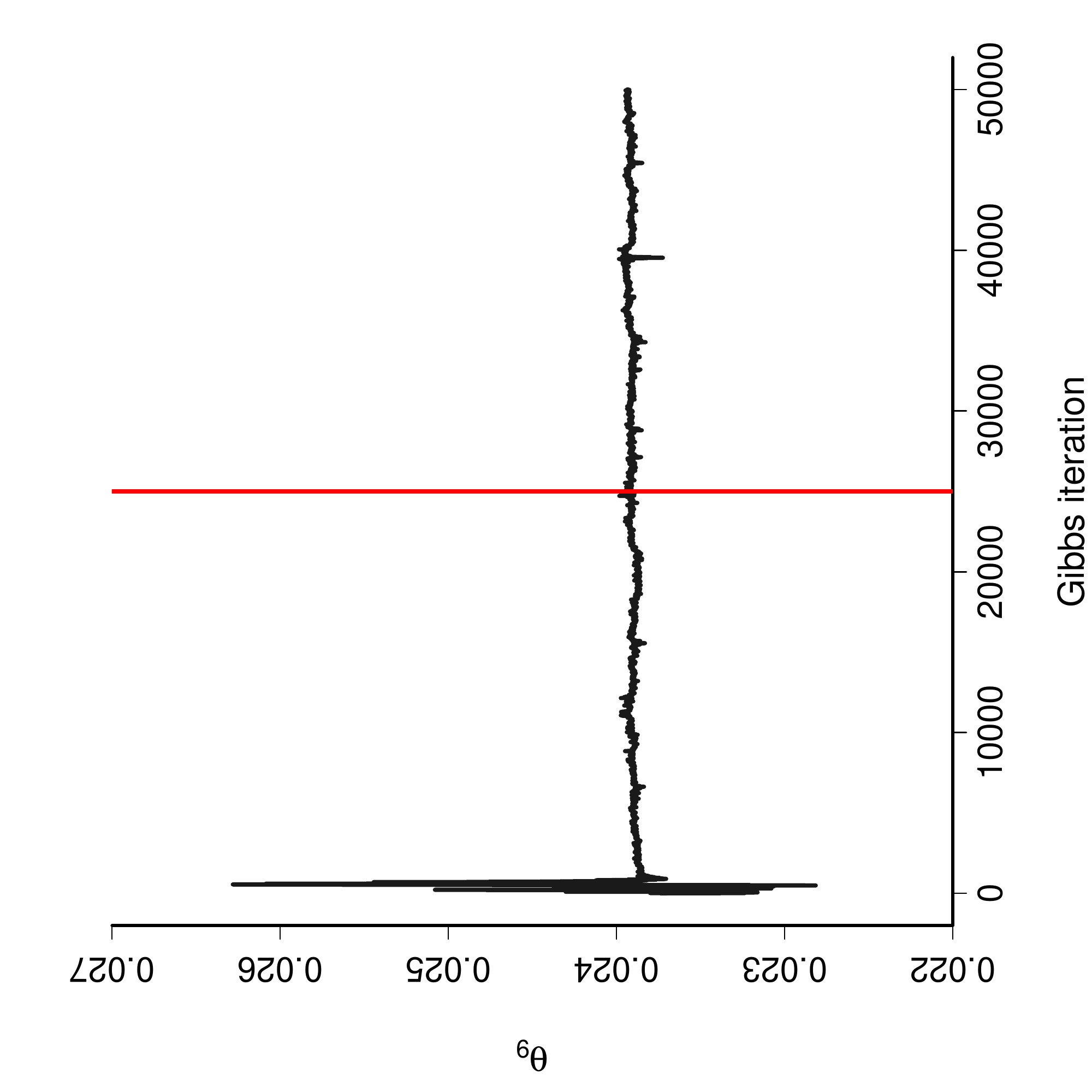}}
	\caption{\small Estimated components of 
	$A_t$. (a) 
	Posterior mean (red line) of the deseasonalised parameter $\exp(\theta_{1, t}^{[1]} + \theta_{3, t}^{[1]} \delta(t))$, with $95\%$ HPD intervals (shading) and L-moments estimates (grey lines); (b) Associated estimates of $\exp(A_t)$ (dots); (c) Estimated seasonal effect of the linear predictor $\lambda_{1,t}$ given the posterior mean for $\theta_{3, t}^{[1]}$; (d) Residual plot for $A_t$, showing the differences $\sobs_{1, t} - \hat{\bflambda}_{1, t}$. Panels (e), (f): sampler trace plots for  $A$ at time $t=1$ and its average summer effect $\bftheta_{9,t}^{[1]}$.
}
	\label{fig:results}
      \end{figure}
   
   The points in Figure \ref{fig:means} are the estimated location parameter means when including the estimated seasonality
   (Figure \ref{fig:Seasonal_effect}), for example, showing an average price increase of around $5.6\%$ from Thursdays to Fridays.
   The residual plot (Figure \ref{fig:Residuals}) exhibits a slight lack-of-fit, suggesting some kind of annual sinusoidal modelling is required. The highest residual was observed on Valentine's weekend when, perhaps, there may be an increase in demand from  couples.
   The lowest residual was on the first day of the high season: Friday, April 1st.

These results were based on  $M=1$ million approximate Gibbs sampler iterations, retaining every 20th sample, and then discarding the first 25,000 iterations as burn-in.  The sampler was initialised from estimates obtained by fitting a simple state space model (that assumes each series in Figure \ref{fig:observed} follows an independent dynamic linear model, with pre-specified matrices $\bfW^{[i]}$) by Kalman smoothing. There are 13,140 unknown parameters in the model, and assessing  chain convergence is not trivial. Trace plots of the location parameter $A$ at time $t=1$ and its average summer effect ($\bftheta_{9,t}^{[1]}$) are displayed in Figure \ref{fig:results}e,f.

\

  It would be extremely challenging for regular ABC methods to handle a model of this size and complexity.
However, computationally this analysis was still expensive -- it took almost 10 days to generate the 1 million Gibbs sampler iterations in {\em R} on a HP device with an Intel Core i7-4790 CPU (3.6GHz) with
16 GB of RAM. In addition, Gibbs samplers result in slowly mixing chains when performing low dimensional parameter block updates, although this low dimensionality is exactly the feature required for ABC methods to function well.
   In this analysis use of Linear Bayes allowed us to model the vector $\bflambda$ jointly, rather than separately for each of its elements. This accounts for its full correlation structure and naturally handles heteroscedasticity. With separate univariate regressions, one would have to accommodate possible interaction terms and model the variance explicitly.

  %%%%%%%%%%%%%%%%%%%%%%%%%%%%%%%
  %%%%%%%%%%%%%%%%%%%%%%%%%%%%%%%
  \section{Discussion} \label{conclusion}
  %%%%%%%%%%%%%%%%%%%%%%%%%%%%%%%
  %%%%%%%%%%%%%%%%%%%%%%%%%%%%%%%
  
  Because it suffers from the curse of dimensionality, ABC performs most effectively for lower dimensional models with lower dimensional summary statistics. In order to consider more complex and higher-dimensional models, such as the 13140 parameter dynamic model considered in Section \ref{application}, this dimensionality must be structurally lowered. This is achieved with the likelihood-free approximate Gibbs sampler. As the full conditional distributions are approximated by regression models, this approach can substantially outperform related Metropolis-Hastings based samplers (e.g.~\shortciteNP{Kousathanas2016}).

  We considered various strategies for constructing the regression models. Localising bespoke regression models at each iteration of the approximate Gibbs sampler can approximate the true conditional distributions more accurately than global regression models that are fitted once, which ultimately leads to lower posterior approximation errors. However, they are correspondingly more expensive to implement. 
  Similarly, simple regression models are faster to fit than more sophisticated models, at the price of greater approximation. The simulations in Section \ref{hierarchical} demonstrated that non-linear deep learning models substantially improved the posterior estimates.

  Similar to the Metropolis-Hastings ABC-MCMC algorithm of \shortciteN{Kousathanas2016}, the likelihood-free approximate Gibbs sampler embraces the spirit of Bayesian modelling with potentially inconsistent conditional distributions, as advocated by \citeN{gelman04}. This potential inconsistency can be greatly diminished  if the fitted regression models
  are sufficiently flexible so that they can approximate the true conditional distributions arbitrarily well. Whether this is possible or not is model and regression model specific. Very recent work by \shortciteN{clarte+rrs19} provides interesting theoretical insights on the conditions under which this will be possible in the ABC context.

  One possible drawback of the likelihood-free Gibbs sampler is that it trades off the greater accuracy of  lower-dimensional ABC models for slower mixing Markov chains, particularly in more complex models, due to the Gibbs updates. However, this is a genuine tradeoff, and for some problems these tools are potentially the only feasible option.
  
.

  %%%%%%%%%%%%%%%%%
  %%%%%%%%%%%%%%%%%%
  \subsection*{Acknowledgements}%% 
  %%%%%%%%%%%%%%%%%%
  %%%%%%%%%%%%%%%%%

  GSR is funded by the CAPES Foundation via the Science Without Borders program (BEX 0974/13-7).
  DJN is supported by a Singapore Ministry of Education Academic Research Fund Tier 1 grant (R-155-000-189-114).
  SAS is supported by the Australia Research Council through the Discovery Project Scheme (FT170100079), and the Australian Centre of Excellence for Mathematical and Statistical Frontiers (ACEMS, CE140100049).
  The authors are grateful to Wilson Ye Chen and Gareth W.~Peters for generously providing the code used to compute the L-moment estimate of parameters of the $g$-and-$k$ distribution.

\bibliography{Approximate_gibbs_sampling}

\section*{Appendix}

\subsection*{A.1: Specification of $\bfF_t$ and $\bfG_t$}

Each $g$-and-$k$ parameter $\bfbeta_{t}^{[1]}$ (with $\bfbeta_t=(\bfbeta_{t}^{[1]},\ldots,\bfbeta_{t}^{[4]})^\top$), $i=1, \ldots, 4$, is defined by its own system parameters, $\bftheta_{t}^{[i]}$, and the matrices 
   $\bfF_{t}^{[i]} = (\bfE_2, \bfE_6, \delta(t))^\top$
  and 
   {\small 
   \begin{align*}
    \bfG_{t}^{[i]}  =\bfG^{[i]} =
      \begin{pmatrix}
        \bfJ_2 & {\bf 0}_{2 \times 6} & {\bf 0}_{2 \times 1} \\
        {\bf 0}_{6 \times 2} & \bfP_6 & {\bf 0}_{6 \times 1} \\
        {\bf 0}_{1 \times 2} & {\bf 0}_{1 \times 6} & 1
      \end{pmatrix}, 
    \quad \text{where } \bfJ_2 =
      \begin{pmatrix}
        1 & 1 \\
        0 & 1 \\
      \end{pmatrix},
    \quad\bfP_6 =
      \begin{pmatrix}
        -{\bf 1}_{1 \times 5} & -1 \\
        {\bf I}_5 & {\bf 0}_{5 \times 1} \\
      \end{pmatrix},
    \end{align*}
    }
   $\bfE_n=(1, 0, \ldots, 0)$ is an $n$-dimensional vector, $\delta(t)$ is an indicator function that takes value $1$ if $t$ is in the summer season and $0$ otherwise, and ${\bf 1}$ denotes a matrix of ones. $\bfJ_2$, which is a Jordan block, implies a local-linear trend for the latent level $\theta_{1, t}^{[i]}$. $\bfP_6$ is a permutation matrix that models the weekly seasonal effect, which impacts the series though $\theta_{3, t}^{[i]}$. The summer-effect is described by $\theta_{9, t}^{[i]}$.  
   The model \eqref{eq.model} becomes fully specified by setting 
   \[
    \bfF_t = \bfF_t^{[i]} \otimes {\bf I_4}, \quad \bfG_t = \bfG^{[i]} \otimes {\bf I_4} \quad \text{and} \quad\bftheta_t=(\bftheta_t^{[1]}, \ldots, \bftheta_t^{[4]}),
   \]
   where $\otimes$ is the Kronecker product.
   This specification imposes those features perceived to drive the {\em Airbnb} data, however alternative models could be adopted.
   For more details on how to specify the matrix of a dynamic model, see e.g.~\shortciteN{Petris2009}.

\subsection*{A.2: Full conditional distributions}

The full conditional distribution (FCD) of the system's initial state $\bftheta_0$ is
%  \[
   $p(\bftheta_0|\cdot) \sim N(\bfa_0, \bfSigma_0),$ 
%  \]
  where $\bfSigma_0 = (\bfG^\top_{1} \bfW^{-1} \bfG_{1} + \bfC_0^{-1})^{-1}$ and $\bfa_0=\bfSigma_0 (\bfC_0^{-1} \bfm_0 + \bfG^\top_{1} \bfW^{-1} \bftheta_1)$.

  To facilitate sampling the system's state $\bftheta_T$, we augment the parameter space to keep track of the parameter $\bftheta_{T+1}$, with FCD given by
   $p(\bftheta_{T+1}|\cdot) \sim N(\bfG_{T+1} \bftheta_T, \bfW)$.

  The FCD of the error's precisions $\tau_i$ are given by
  \[
   p(\tau_i|\cdot) \sim \text{Gamma}\left( \alpha+\frac{T+1}{2}, \nu+ \frac{\sum_{t=1}^{T+1} \bfw_{ti}^2}{2} \right),
  \]
  where $\bfw_t = \bftheta_t - \bfG_t \bftheta_{t-1}$ represents the system innovation at time $t$.

  For the system state $\bftheta_t$, 
  the model equations imply that  
   {\small\begin{align*}
       \begin{pmatrix}
         \left.
         \begin{array}{c}
         \bftheta_t \\
         \bflambda_t
         \end{array}
         \right|
       & \hspace{-.2cm} \bftheta_{t-1}, \bftheta_{t+1}, \bfW
    \end{pmatrix}
    \sim \text{N}
    \begin{bmatrix}
      \begin{pmatrix}
        \bfa_t \\
        \bff_t
      \end{pmatrix},
      &
      \begin{pmatrix}
        \bfR_t & \bfR_t F_t  \\
        F^\top_t \bfR_t & \bfq_t 
      \end{pmatrix}
    \end{bmatrix}, 
    \end{align*}}
    where 
      $\bff_t = \bfF^\top_t \bfa_t,$ 
      $\bfq_t = F^\top_t \bfR_t F_t,$
       $\bfa_t = \bfR_t (\bfW^{-1} \bfG_t \bftheta_{t-1} + \bfG^\top_{t+1} \bfW^{-1} \bftheta_{t+1}),$ 
      and
      $\bfR_t = (\bfG^\top_{t+1} \bfW^{-1} \bfG_{t+1} + \bfW^{-1})^{-1}.$
   It then follows from the conditional properties of the multivariate normal distribution that 
    $p(\bftheta_t | \bftheta_{t-1}, \bftheta_{t+1}, \bfW, \bflambda_t) = N(\bfmu_t, \bfSigma_t),$
   where
   $\bfmu_t=\bfa_t + \bfR_t \bfF_t \bfq_t^{-1} (\bflambda_t - \bff_t)$
   and 
    $\bfSigma_t=\bfR_t - \bfR_t \bfF_t \bfq_t^{-1} \bfF^\top_t \bfR_t.$

\end{document}